\documentclass{aa}  
\usepackage{graphicx}
\usepackage{url}
\usepackage{tabularx}
\usepackage[figuresright]{rotating}
\usepackage{txfonts}
\newcommand\srm{\scriptscriptstyle\rm}
\def\lll{$\lambda\lambda$}
\def\ll{$\lambda$}
\def\halphal{H\,$\alpha$\,$\lambda$6562}
\def\halpha{H\,$\alpha$}
\def\hgamma{H\,$\gamma$}
\def\hgammal{H\,$\gamma$\,$\lambda$4340}
\def\hbetal{\,H$\beta$\,$\lambda$4861}
\def\hbeta{H\,$\beta$}
\def\OIII{[O\,${\srm III}$]}
\def\OIIId{[O\,${\srm III}$]\,$\lambda\lambda$4959, 5007}
\def\MgII{Mg\,${\srm II}$}

\def\MgIIl{Mg\,${\srm II}$\,$\lambda$2798}
\def\NeV{[Ne\,${\srm V}$]}
\def\NeVb{[Ne\,${\srm V}$]\,$\lambda$3425}
\def\NeIV{[Ne\,${\srm IV}$]}
\def\OII{[O\,${\srm II}$]}
\def\OIIl{[O\,${\srm II}$]\,$\lambda$3727}
\def\FeII{Fe\,${\srm II}$}
\def\NIId{[N\,${\srm II}$]\,$\lambda\lambda$6548, 6583}

\def\SIId{[S\,${\srm II}$]\,$\lambda\lambda$6716, 6731}
  
\begin{document}

   \title{Multi-wavelength study of the gravitational lens system RXS~J1131-1231}

   \subtitle{III. Long slit spectroscopy: micro-lensing probes the QSO structure
    \thanks{Based on observations collected at the European Southern
    Observatory, Paranal, Chile (ESO Program 71.A-0407(B, E))}}

  \author{D. Sluse\inst{1,2}, J.-F. Claeskens\inst{2},
    D. Hutsem\'ekers\inst{2}\thanks{Ma\^itre de recherches du
      F.N.R.S. (Belgique)} \and J. Surdej\inst{2}\thanks{Directeur de
      recherches honoraire du F.N.R.S. (Belgique)} }

   \offprints{dominique.sluse@epfl.ch}

   \institute{ Laboratoire d'Astrophysique, Ecole Polytechnique F\'ed\'erale de Lausanne
    (EPFL) Observatoire, 1290 Sauverny, Switzerland \and Institut d'Astrophysique et de G\'eophysique, Universit\'e de Li\`ege,
    All\'ee du 6 Ao\^ut 17, B5C, B-4000 Sart Tilman, Belgium
             }

   \date{Received 28/11/2006; Accepted 27/02/2007}

 
  \abstract
   {}
   {We discuss and characterize micro-lensing among the 3 brightest lensed images (A-B-C) of the gravitational lens system RXS J1131-1231 (a quadruply imaged AGN) by means of long slit optical and NIR spectroscopy. Qualitative constraints on the size of different emission regions are derived. We also perform a spectroscopic study of two field galaxies located within 1.6 arcmin radius from the lens. }
   { We decompose the spectra into their individual emission components using a multi-component fitting approach. A complementary decomposition of the spectra enables us to isolate the macro-lensed fraction of the spectra independently of any spectral modelling. }
   {1. The data support micro-lensing de-amplification of images A \& C. Not only is the continuum emission microlensed in those images but also a fraction of the Broad Line emitting Region (BLR).\\
   2. Micro-lensing of a very broad component of \MgII~emission line suggests that the corresponding emission occurs in a region more compact than the other components of the emission line. \\
   3. We find evidence that a large fraction of the \FeII~emission arises in the outer parts of the BLR. We also find a very compact emitting region in the ranges 3080-3540\,\AA~ and 4630-4800\,\AA~that is likely associated with \FeII. \\
   4. The \OIII~narrow emission line regions are partly spatially resolved. This enables us to put a lower limit of $\sim 110h^{-1}$\,pc on their intrinsic size. \\
   5. Analysis of \MgII~absorption found in the spectra indicates that the absorbing medium is intrinsic to the quasar, has a covering factor of 20\%, and is constituted of small clouds homogeneously distributed in front of the continuum and BLRs. \\
   6. Two neighbour galaxies are detected at redshifts $z=0.10$ and $z=0.289$. These galaxies are possible members of galaxy groups reported at those redshifts. 
}
   {}

    \titlerunning{Long slit spectroscopy of RXS
    J113155.4-123155.}  \authorrunning{Sluse D. et al.}


   \maketitle
%

\section{Introduction}
\label{sec:intro}

  \begin{figure}
  \begin{center}
  \includegraphics[width=7.7cm]{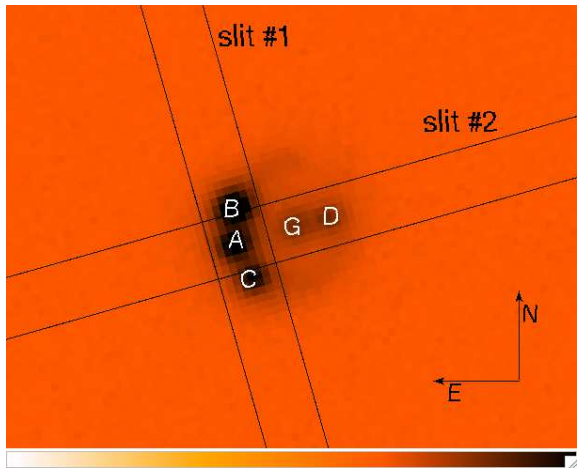}
       \end{center}
       \caption{Image of J1131 (0.8\arcsec seeing) with superimposed
         slit orientations \#1 and \#2. Identification of the quasar
         lensed images (A, B, C, D) and of the lensing galaxy (G) are
         also reported. }
       \label{fig:slit}
   \end{figure}

RXS J113155.4-123155 (hereafter J1131) is one of the nearest confirmed
multiply imaged AGN. The source at $z_s = 0.66$ is lensed by an
elliptical galaxy at $z_l = 0.295$ (Sluse et al. \cite{SLU03}). This
system is a long axis quad with an image configuration very similar to
B1422+231 (Patnaik et al. \cite{PAT92}): three merging images (B-A-C;
typical of a source lying close to a cusp caustic) face the faint
saddle-point image D lying close to the lensing galaxy G
(Fig.~\ref{fig:slit}). 

J1131 has several observational advantages compared to the
other lens systems, namely an integrated magnitude R $\sim$ 15.3 and a
large angular separation between the lensed images
($\Delta\theta \sim$ 3.6$\arcsec$). These characteristics make J1131 an
ideal target for spectroscopic studies (e.g. Sheinis~\cite{SHE06}) and
optical flux monitoring (e.g. Morgan et al.~\cite{MOR06}; COSMOGRAIL
project). Additionally, the low redshift of the source quasar enables
one to observe the \MgII, \hbeta~and \OIIId~emission lines in the
optical range. This allows the comparison of the flux ratios in three
different emitting regions, namely the AGN continuum, the broad
emission lines (e.g. \hbeta) and the narrow emission lines
(e.g. \OIIId). Because of their different angular sizes, these three
emitting regions will not be similarly affected by micro-lensing due
to stars in the lensing galaxy or milli-lensing by more massive
substructures (typically $M>$ 10$^4 M_{\odot}$). Therefore, the
comparison of the flux ratio measured in those three regions may offer
a robust diagnostic of micro/milli-lensing at work in the system.

Although milli-lensing had been early suspected in J1131 (e.g. Keeton
et al.~\cite{KEE03b}), Sluse et al. (\cite{SLU06}, Paper I) gave
stronger evidence for the presence of micro-lensing, on the basis of a
multi-wavelength study of the flux ratios observed at different
epochs. Even when tentatively corrected for micro-lensing, the
observed flux ratios were still impossible to reproduce with simple
smooth lens models. Claeskens et al. (\cite{CLA06}) showed in Paper II
that multipole-like models could formally solve the anomaly. For this
purpose, they constructed lens models constrained by the relative
astrometry of the QSO images provided by the HST images and by
substructures present in the nearly complete Einstein ring . This also
enabled them to reconstruct the host spiral galaxy of the source AGN
and to study in detail its properties. In the present paper, we take
advantage of long slit spectroscopy of J1131 to further study
micro-lensing in this system and phenomenologically describe its
effect on different emitting regions and in particular on the broad
line emitting region (BLR).

Abajas et al. (\cite{ABA02}) have shown, using the Kaspi et
al. (\cite{KAS00}) relation linking the BLR size to the quasar
luminosity, that the BLR of {\it less luminous} quasars is likely
small enough (i.e. a few micro-arcseconds) to be affected by
micro-lensing. Because the source in J1131 is rather a (bright)
Seyfert than a luminous quasar, one may thus expect to observe
differential microlensing of its BLR. Thanks to the large spectral
coverage of our data (i.e. $2400< \lambda < 6900$\,\AA~; rest
frame) we looked for micro-lensing of the BLR and compared its effect
on the main Balmer lines and on the \MgII~line. Additionally, we
compared the flux ratios in these lines to those in several narrow
emission lines (NELs), namely \NeV, \OII~and \OIII. Finally, we
studied micro-lensing of the blended \FeII~emission all along the
spectra. The location of the \FeII~emitting region is still hotly
debated (e.g. Zhang et al.~\cite{ZHA06b}) and micro-lensing of the
latter offers the opportunity to compare the size of the
\FeII~emitting region with the BLR size.

The structure of the paper is the following. We describe the
observations and the spectrum extraction in Sect.~\ref{sec:obs}. We
perform a preliminary analysis to explore the data in
Sect.~\ref{sec:explore} and provide more quantitative measurements of
the spectral differences by means of a multicomponent spectral
decomposition in Sect.~\ref{sec:multicomp}. We propose a comprehensive
micro-lensing scenario in Sect.~\ref{sec:discussion} and discuss the
implications of the differential micro-lensing of the broad emission
lines (BELs) in Sect.~\ref{sec:probe}. We present some additional
results concerning the extended nature of the NELs, the absorption
systems observed in the QSO spectra in Sect.~\ref{sec:complementary}
together with the redshifts of the lens and of two neighbour
galaxies. We comment on some working assumptions in
Sect.~\ref{sec:caveats} and summarize the results in
Sect.~\ref{sec:conclusions}.

Unless explicitly stated, we adopt $H_0=$ 70\,km\,
s$^{-1}$\,Mpc$^{-1}$, $\Omega_0=$ 0.3 and $\Lambda_0=$ 0.7; all the
$\chi^2$ values reported are reduced values.

 \section{Observational overview} 
 \label{sec:obs}
 
\subsection{Data}    
\label{subsec:data}

We obtained long slit spectra of J1131 with the FORS2 instrument
mounted on the Kueyen (UT2) ESO Very Large Telescope on April 26th
2003. These data consist of two series of spectra with the 1$\arcsec$
slit oriented along the J1131 lensed images B-A-C and A-D respectively
(hereafter orientations \#1 and \#2; see Fig~\ref{fig:slit}). The
Longitudinal Atmospheric Dispersion Corrector (Avila et
al. ~\cite{AVI97}) has enabled us to keep the slit centering within
$\sim$ 0.2\arcsec all along the wavelength range, reducing the slit
losses in the blue range. The grisms GR600B and GR600I+OG590 blocking
filter have been used, allowing to cover the wavelength ranges $3890
< \lambda < 6280$\,\AA~and $6760 < \lambda < 8810$\,\AA~ with a
resolving power around 800 and 1500 at the central wavelength. The CCD
camera (SR collimator) was unbinned (0.063$\arcsec$/pix in the spatial
direction) in order to ease the spatial deblending of the
spectra. Long slit spectra (with slit orientations identical to FORS
spectra) have also been obtained in the near infrared range ($9810 <
\lambda < 11390$\,\AA) with the ISAAC instrument placed at the focus
of the VLT UT1 telescope (Antu), 13 days before the FORS spectra. The
low resolution grating combined with a 1$\arcsec$ slit and a SZ filter
leads to a resolving power of 550. Table~\ref{tab:log} gives a short
log of the observations.

\begin{table}[]
\caption{Log of the observations for slit orientations \#1 and \#2 (Fig~\ref{fig:slit}). Col.~1: Observing date; Col.~2: Instrument; Col.~3: Total exposure time; Col.~4: Mean seeing during the observations; Col.~5: Mean airmass {\bf sec($z$)}.}
\label{tab:log}
\centering
\begin{tabular}{lcccc}
\hline \hline
Date & Instrument & Exp & seeing & sec($z$) \\
(dd-mm-yy) &  & (s) & (\arcsec) & \\
\hline 
\multicolumn{5}{c}{slit \#1}\\
\hline
26-04-03 & FORS2 (GR600I) & 4 $\times$ 240 & 0.55 & 1.33 \\ 
26-04-03 & FORS2 (GR600B) & 4 $\times$ 240 & 0.4-0.7 & 1.45 \\
13-04-03 & ISAAC (LR+SZ) & 6 $\times$ 200 & 0.45-0.62 & 1.19\\
\hline
\multicolumn{5}{c}{slit \#2}\\
\hline
26-04-03 & FORS2 (GR600I) & 3 $\times$ 430 & 0.53 & 1.07\\
26-04-03 & FORS2 (GR600B) & 3 $\times$ 430 & 0.6-0.8 & 1.12\\
13-04-03 & ISAAC (LR+SZ) & 14 $\times$ 200 & 0.56-0.89 & 1.40\\
\hline
\end{tabular} 
\end{table}
 
\subsection{Reduction}

The FORS spectra were bias subtracted and flatfielded using
IRAF{\footnote{IRAF is distributed by the National Optical Astronomy
    Observatories (NOAO), which are operated by the AURA, Inc., under
    cooperative agreement with the National Science Foundation
    (NSF).}}. A set of 5 dome flat fields were used. The wavelength
dependent structure of the flat fields was corrected by fitting a high
order cubic spline with the {\texttt{response}} task. The resulting
flat field is of sufficient quality to correct the fringing
appearing on the raw data. The wavelength calibration was performed
thanks to He-Hg-Cd and He-Ar-Ne calibration arcs. Since images B and C
are not well centered in the slit, we shifted the zeropoint of
their wavelength calibrated spectra by $\sim$ 0.8\,\AA~with respect to
A. Finally, the sky background was removed by fitting and subtracting
a first order Chebychshev polynomial in the spatial direction in two
adjacent regions of 180 pixels width not illuminated by the
object. Since no standard star was observed during the science
night, the sensitivity curve was constructed using the two
standard stars GD108 (for the GR600B grism) and LTT7379 (for the
GR600I grism) observed on April 15th 2003 and May 25th 2003,
respectively, through a 2.5$\arcsec$ width slit. The same reduction
procedure was applied to these stars, using calibration data
associated with these observing dates. The stars and the objects were
corrected for sky extinction with the Paranal extinction estimates
available on the ESO website.
 
For the ISAAC spectra, as is standard practice in the infrared, the
object was observed in two positions along the slit. The strong and
highly variable night sky emission lines were effectively removed by
subtracting the resulting spectra from each other. The 2-D
sky-subtracted spectra were then flat-fielded, registered, wavelength
calibrated and co-added. The wavelength calibration was performed
using both xenon argon arc lines and OH sky lines (Rousselot et
al. ~\cite{ROU00}). The slit curvature correction was applied using
the {\texttt{fitcoords}} and {\texttt{transform}} IRAF tasks. Once
corrected for the slit curvature, the NIR spectra were still inclined
with respect to the pixel grid. We thus applied an additional rotation
of 0.78$\degr$ to align the spectra with the pixel grid and ease the
subsequent spectrum extraction (Sect.~\ref{subsec:extract}). Finally,
the response curve was calculated by dividing the reduced spectrum of
the B2 type standard star HIP88126 by a black body at 20000\,K. For
both optical and NIR data, a noise image was constructed at each step
of the reduction, using the standard error propagation formula.
 
\subsection{Spectrum extraction}
\label{subsec:extract}

\begin{figure}
\centering
\includegraphics[width=0.95\columnwidth]{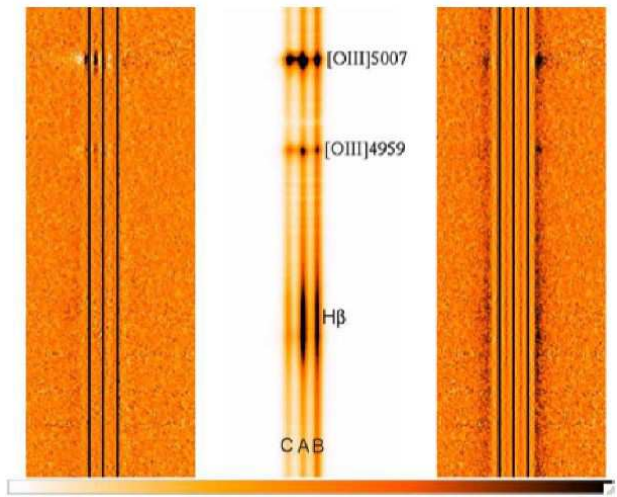}
\caption{ Central panel : Bidimensional Moffat models of the spectra of C, A \& B in the \hbetal~\& \OIIId~ region. Left and right panels: bidimensional residuals ($\pm$3$\sigma$) for a Moffat and a Gaussian fit. Vertical black lines indicate the center of the profiles for C, A, B.}
\label{fig:2Dfit}
\end{figure}

\begin{figure*}
\centering
\includegraphics[width=0.65\textwidth, angle=-90]{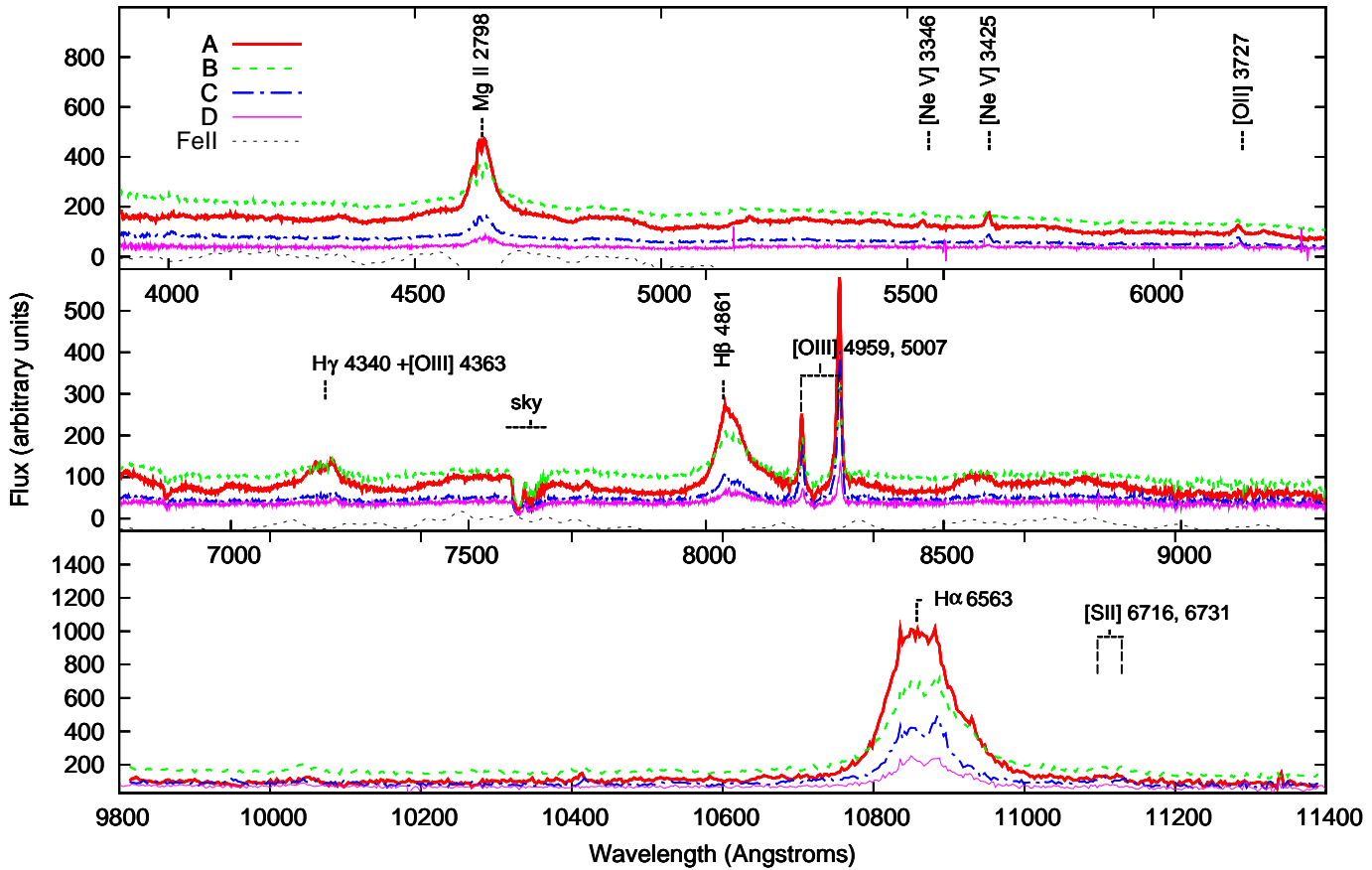}

\caption{Extracted spectra of images A (thick solid red), B (dashed green), C (dashed-dotted blue) and D (thin solid pink) in the blue (up), red (middle) and NIR (bottom) ranges. Black dashed line (bottom of the graph) in the blue and red ranges is the template of \FeII~emission redshifted at $z=0.657$  and arbitrarily shifted in flux for legibility. Relative flux ratios between the spectra are conserved except for image D. Main emission lines are identified. }
\label{fig:spectra}
\end{figure*}

Our extraction method is different for slit orientations \#1 and
\#2. For frames obtained with slit orientation \#1, only the
point-like images A, B and C were in the slit. We thus fitted three
Moffat profiles in the spatial direction for each wavelength bin
independently, allowing both the centroid of the Moffat, the slope
$\beta$ and $FWHM$ to vary. After this preliminary fit of the
bi-dimensional spectra, we smoothed the fitted centroid positions of
A-B-C with a 10\,\AA~moving average and reprocessed the extraction by
fixing the positions to the ones found after smoothing. The fitted 2D
spectra were then compared to the observed ones in order to track for
extraction artifacts. Figure~\ref{fig:2Dfit} compares the residuals
observed in the \hbeta~and \OIII~region (i.e. $7900 < \lambda <
8350$\,\AA) when extraction is performed with 3 Moffat (left panel)
and with 3 Gaussian (right panel) profiles. One can see systematic
residuals next to each lensed image spectrum when Gaussian profiles are
used. These residuals are indeed observed all along the spectra. On
the other hand, the residuals are flat at all wavelengths when Moffat
profiles are used, except around \OIIId. In order to explain the
origin of those residuals we constructed a 2D pseudo-continuum under
\OIIId~by interpolating the continuum level measured just before and
after \OIII. Once that pseudo-continuum had been subtracted from the
data, we obtained 2D spectra of \OIII~emission only. The fit of the
latter with 3 Moffat profiles is significantly worse than the fit of
the same region for the continuum only ($\chi^2=1.01$ for the
continuum and $\chi^2=3.71$ for \OIIId). The results are even worse if
we impose the Moffat parameters for \OIII~ to be equal to those found
for the continuum ($\chi^2=4.68$). This is a hint that the observed
\OIIId~emission is not point-like but is partially resolved. The
complete extracted spectra of A, B and C are displayed in
Fig.~\ref{fig:spectra}. The flux ``lost'' in the \OIII~region due to
imperfect extraction is only 0.5\% of the \OIIId~emission in
A+B+C. Therefore, our extraction method does not introduce significant
error on the flux measurement in \OIII~due to its spatial extension.

After extraction, we constructed a synthetic optical spectrum (sampled
on a pixel grid of 1.33\,\AA) of each lensed image by joining the blue
(GR600B) and red (GR600I) spectra on a common grid. Since there is no
overlapping domain between the blue and red spectra, we assume that
the fractional loss of flux is similar in each spectrum such that no
relative flux rescaling between blue and red is necessary. We did not
attempt to rescale in flux the NIR spectra over the optical ones since
these spectra were obtained 13 days later. Although A, B and C
are not perfectly aligned along the slit, we assume that the relative
flux loss between the lensed images is small. To test this, we
extracted the flux in the acquisition image along 3 different
rectangular apertures of 0.8, 1.0 and 1.2$\arcsec$ widths, each
oriented like the true slit. We then fitted A, B, C with 3
one-dimensional Moffat profiles as we did for the spectrum
extraction. Following this procedure we found that the measured flux
ratios are independent of the chosen slit width.  Additionally, these
flux ratios agree within 5\% with the ratios of the spectra multiplied
by the $R$-band transmission curve.

For the slit orientation \#2, we performed classical extractions in
apertures centered on the lensed images A and D and around G, using
the IRAF task \texttt{apall}. Unfortunately, the extracted spectra of
D were significantly contaminated by the lensing galaxy while the one
of A was likely contaminated by images B and C. Although we could
adapt the size of the extraction aperture to reduce the contamination,
this precluded a correct flux calibration for the spectrum of D. For
this reason, we do not analyse that image in the
remaining. Additionally, three other objects are located in the
slit. A star (11:31:55.8, -12:32:16) and two galaxies {\it{gal}}\#1
and \#2 (11:31:55.025, -12:32:13 and 11:31:57.7, -12:32:22) located at
55 and 95\arcsec~from the lens. They were extracted through fixed
apertures of 10 pixel width. These spectra are analysed separately
(Sect.~\ref{subsec:lens}) from those of the lensed images. The latter
are described in the following two sections.

\section{Exploratory analysis}
\label {sec:explore}

The goal of this section is to identify general features in the A-B-C
lensed image spectra, by means of a phenomenological and model
independent approach. Although numerical results will be obtained,
more quantitative results derived from systematic fittings will be
presented in Sect. \ref{sec:multicomp}.

\subsection{Spectrum ratios}

\begin{figure}
\begin{center}
\includegraphics[width=6.cm,angle=-90]{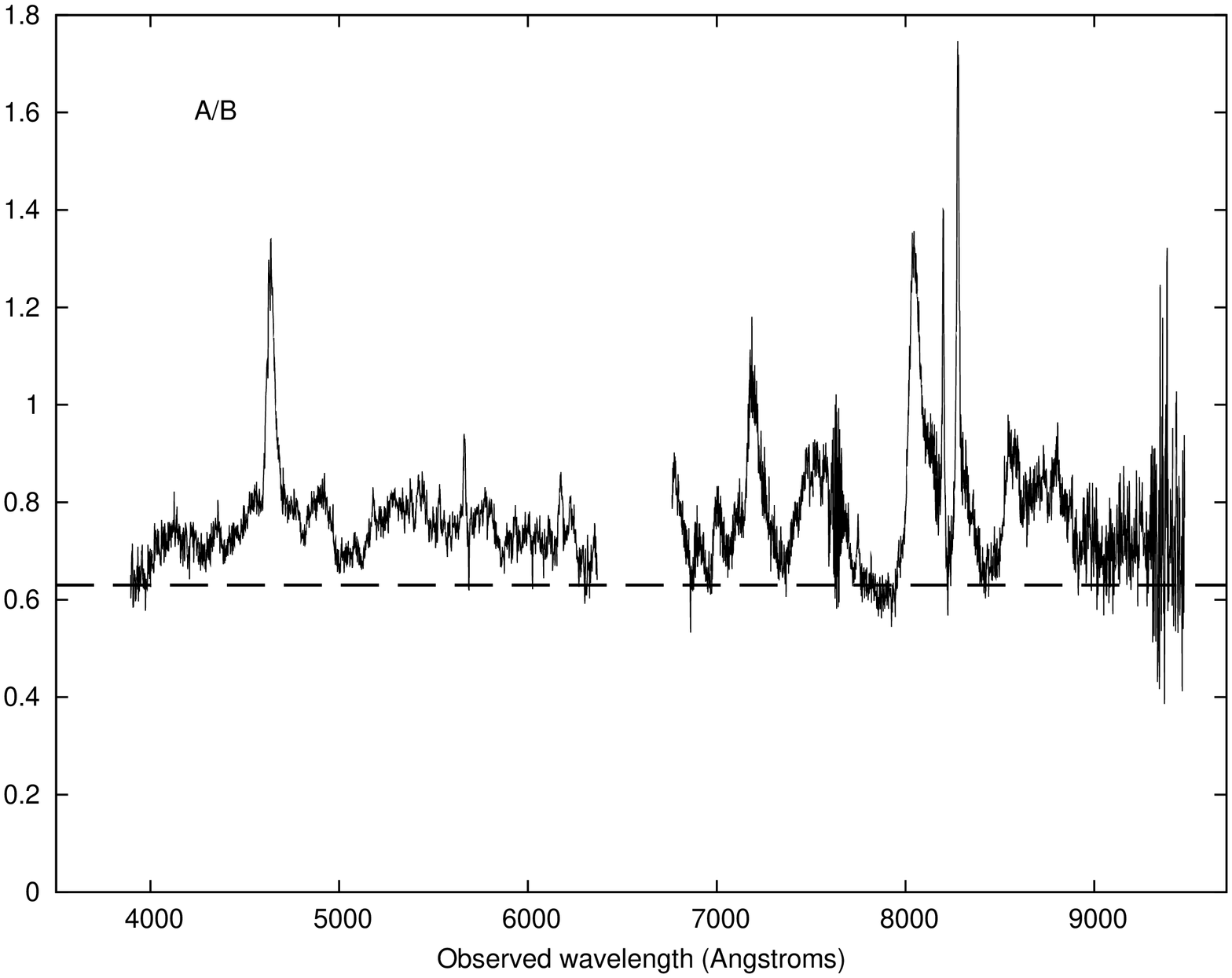}
\includegraphics[width=6.cm,angle=-90]{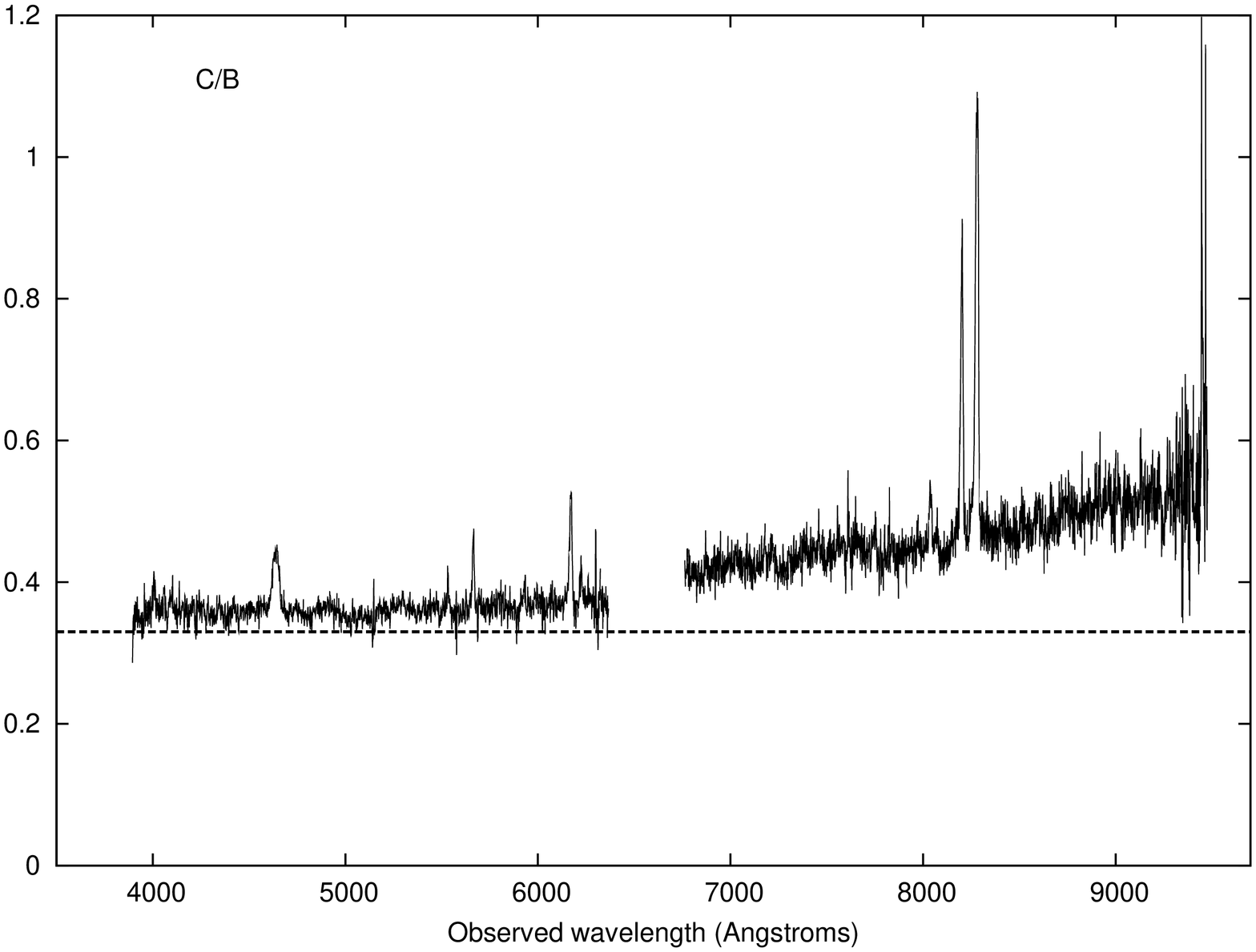}
\caption{Ratio of spectra A \& B (top) and C \& B (bottom). The
  observed (flat) ratio $R=M\mu$ is plotted with a dotted line.  $R=
  0.63$ for the A/B ratio and $R= 0.33$ for the C/B ratio.}
\label{fig:spec_ratio}
\end{center}
\end{figure}

The first insight into the relative differences between the lensed
spectra is simply provided by computing the ratios of individual
spectra. Indeed, since gravitational lensing is achromatic, one
expects the lensed image spectra to be identical up to a magnification
factor. Therefore the spectral ratio between two images is expected to
be flat. However, extinction in the lensing galaxy may differently
affect the lensed images, as well as micro/milli-lensing produced by
substructures in the lens, such that chromatic trends may appear in
the spectral ratios (e.g.  Jean \& Surdej ~\cite{JEA98}, Wambsganss \&
Paczy\'nski~\cite{WAM91}). In some cases, micro/milli-lensing can also
modify the profiles of the BELs (e.g. Abajas et al.~\cite{ABA02}). The
ratio between individual spectra thus provides one with a simple
diagnostic of the presence of differential extinction or
micro-lensing.  The A/B and C/B ratios are presented in
Fig. \ref{fig:spec_ratio}.

One can clearly see the imprint of the QSO emission line spectra in
the A/B spectral ratio, indicating that the flux ratio is different in
the continuum (e.g. 7900-7950~\AA), the BELs (e.g. 8000-8170~\AA) and
the NELs (e.g. 8240-8320~\AA). Differences in the pseudo-continuum
(i.e. the AGN power law continuum+blended \FeII~emission) are also
apparent (e.g. 5100-6200~\AA). This suggests that ``classic''
microlensing is probably at work by affecting the compact region
emitting the continuum. However, no chromatic dependence is observed,
as illustrated by the flat underlying continuum ratio (tentatively
indicated by the dotted line in Fig. \ref{fig:spec_ratio}, at the
value $R=0.63$) 

On the other hand, the C/B spectral ratio is not flat.  First, most of
the BEL structure has disappeared, with the noticeable exception of
the \MgII\, line, in which the flux ratio is significantly different
from the one in the underlying continuum. Second, the flux ratio in
the NELs (in particular in \OIIId) is drastically different from the
one in the continuum (i.e. as measured at the foot of the line
``ratio''). Finally, an overall, {\em chromatic} trend is observed
with a flux ratio increasing from $\sim 0.35$ in the blue to $\sim
0.55$ in the red. An attempt to provide a coherent interpretation with
a minimum number of hypotheses is discussed below.

\subsection{The microlensed spectra}
\label{subsec:exploreFmu}

\subsubsection{The $F\mu$ method}
\label{sec:decomp}

Assuming the observed spectra $F_i$ are simply made of a superposition
of spectrum $F_M$ which is only macro-lensed and of a spectrum
$F_{M\mu}$ both macro and {\it micro}-lensed, it is easy to extract
both components $F_M$ and $F_{M\mu}$ by using pairs of observed
spectra. However, the relative macro amplification $M$ must be known.

Indeed, defining $M=M_1/M_2 \,(>0)$ and $\mu = \mu_1/\mu_2 \,(>0)$ as
the constant macro- and micro-amplification ratios between image
1 and image 2, we have:
 
\begin{equation}
\begin{array}{l}
F_1  = M F_M + M \mu F_{M\mu}\\
F_2  = F_M + F_{M\mu} \,.\\
\end{array}
\label{eq:decomp1}
\end{equation}

\noindent The latter equations can be rewritten to extract $F_M$ and $F_{M\mu}$:

\begin{equation}
\begin{array}{l}
F_M  = \frac{F_1/M - \mu F_2}{1-\mu} \\
F_{M\mu} =  \frac{F_2- F_1/M}{1-\mu}\,, \\
\end{array}
\label{eq:decomp2}
\end{equation}

\noindent
where $\mu$ must be chosen to satisfy the positivity constraint $F_M
> 0$ and $F_{M\mu} >0$.

It is clear from Eq.~(\ref{eq:decomp2}) that $F_{M\mu}$ is directly
determined from the macro-amplification ratio $M$ and the observed
spectra $F_1$ and $F_2$, up to a scaling factor.  On the other hand,
if $\mu$ and $\mu'$ are two values of the micro-amplification ratio
which satisfy the positivity constraints and such that $\mu' < \mu$,
it is easy to show that:

\begin{equation}
F_M(\mu') = F_M(\mu) + \frac{\mu-\mu'}{1-\mu'} F_{M\mu}\,.
\label{eq:FM}
\end{equation}

\noindent It is obvious from the previous equation that adopting the smaller
value $\mu'$ is equivalent to add a fraction of the microlensed
spectrum to the macrolensed spectrum obtained with $\mu$. Since an
emitting region can only be microlensed {\em or} not, the only
possible choice for $\mu$ is the {\it maximum} value satisfying the
positivity constraint.

The previous decomposition is exact when $\mu$ is constant as a
function of wavelength, i.e. when the micro-lensing ratio is
achromatic. However microlensing is a function of the source size
(which is wavelength dependent) and this function can be different for
different images, so that even the micro-lensing {\it ratio} $\mu$ can
be chromatic.  Any such wavelength dependence of $\mu$ would be
propagated into the spectra $F_M$ and $F_{M\mu}$. Although this is not
important as long as the method is simply used to identify spectral
features affected by {\it some} micro-lensing (i.e. whose flux ratio
is not equal to the macro value), we shall show that $\mu=$constant is
a reasonable choice.

In the following, we will call the decomposition of the
spectra into $F_M$-$F_{M\mu}$, the $F\mu$ method.

\begin{figure*}[!t]
\begin{center}
\includegraphics[width=10cm, angle=-90]{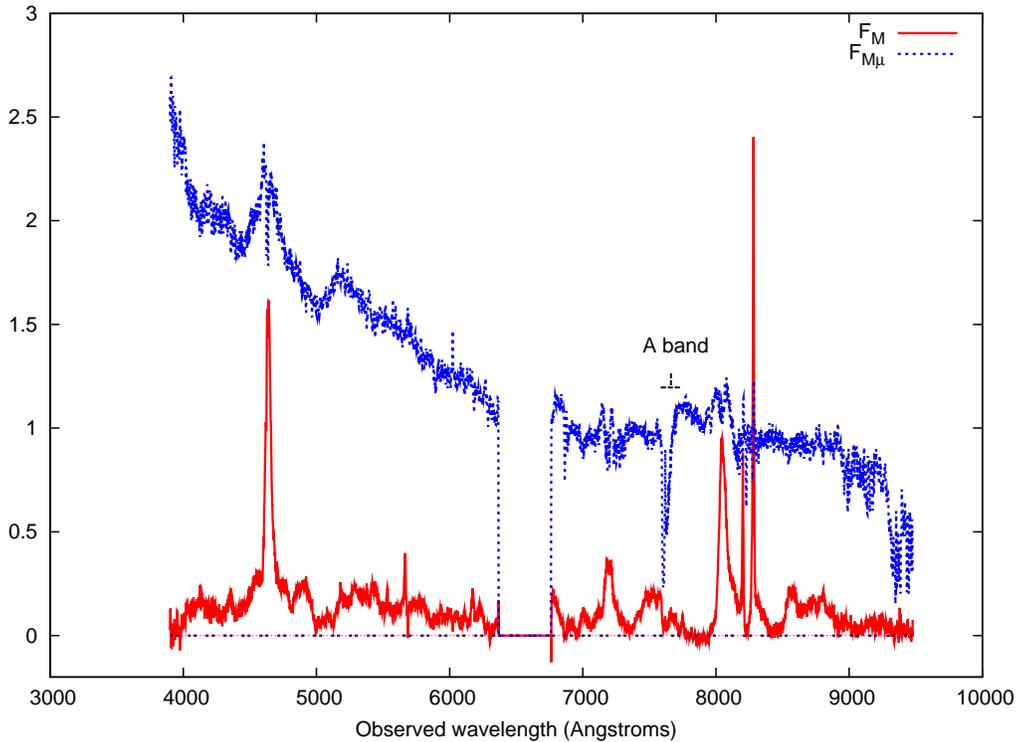}
\caption{Fraction of the spectrum affected ($F_{M\mu}$, dotted) and unaffected ($F_M$, solid) by microlensing  (arbitrary units) in the A-B spectrum pair (see Eq.~(\ref{eq:decomp2})). The A-band telluric absorption is also identified.}
\label{fig:fmfmuAB}
\end{center}
\end{figure*}

\subsubsection{The A-B pair}
\label{sec:ABpair}

In order to apply the $F\mu$ method to the A-B pair of spectra, the
macro amplification ratios must be fixed. In the remainder of the
paper, we use the hypothesis that they are given by the observed flux
ratios in the \OIIId~lines. This seems a reasonable choice in the
sense that the NELs are produced by the largest emitting region, thus
the least affected by microlensing. The fact that the \OIIId~emission
is partially {\em resolved}, especially between images C and A (see
Sect.~\ref{subsec:extract}) does not bias the \OIII~flux
estimate. Indeed, we have shown that only a very small fraction
(i.e. $<$1\%) of the flux in \OIII~was lost during the extraction
(Sect.~\ref{subsec:extract}).

Under this hypothesis, first we must have that $M=M_A/M_B = 2.1$ in
order to suppress the \OIIId~emission lines in the combined spectrum
$F_A - MF_B$. Second, by associating image A (resp. B) with number 1
(resp. 2) in Eqs.~(\ref{eq:decomp2}), we find $\mu = \mu_A/\mu_B =
0.3$. The extracted spectra $F_M$ and $F_{M\mu}$ are shown in
Fig.~\ref{fig:fmfmuAB}. Note that a constant value of $\mu$ is
certainly a good approximation, given the flat A/B underlying
continuum (Fig.~\ref{fig:spec_ratio})

The striking result of this decomposition is that, in addition to the
continuum, the {\em broadest part of the BEL} is also affected by
microlensing. This might indicate that at least part of the BLR is
very compact. A more thorough discussion is delayed to
Sect.~\ref{sec:discussion}, after a more quantitative analysis is made
in Sect.~\ref{sec:multicomp}.

On the other hand, as expected, the unaffected spectrum $F_M$ is
basically flat and consists of the NEL and narrowest part of the
BEL. Indeed, not only the \OIIId~lines, but also the core of \MgII,
\hgamma~and \hbeta~BELs are prominent. Interestingly, blended \FeII\
emission is also clearly observed around \MgII\ and blue/red-ward of
\hbeta.

\subsubsection{The B-C pair}
\label{sec:BCpair}

\begin{figure}
\begin{center}
\includegraphics[width=6.5cm,angle=-90]{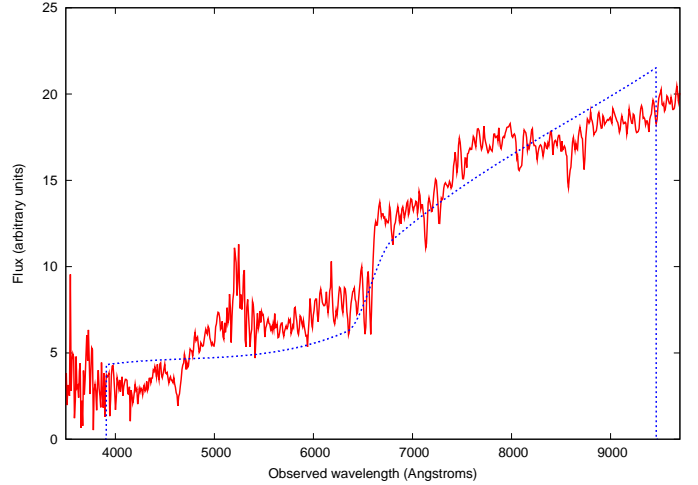}
\caption{The smoothed host galaxy spectrum (dotted) responsible
    for the chromatic trend observed in the C/B spectrum ratio,
    compared with a redshifted spectral template of a Sb galaxy (solid
    ).}
\label{fig:host}
\end{center}
\end{figure}

\begin{figure*}
\begin{center}
\includegraphics[width=10cm, angle=-90]{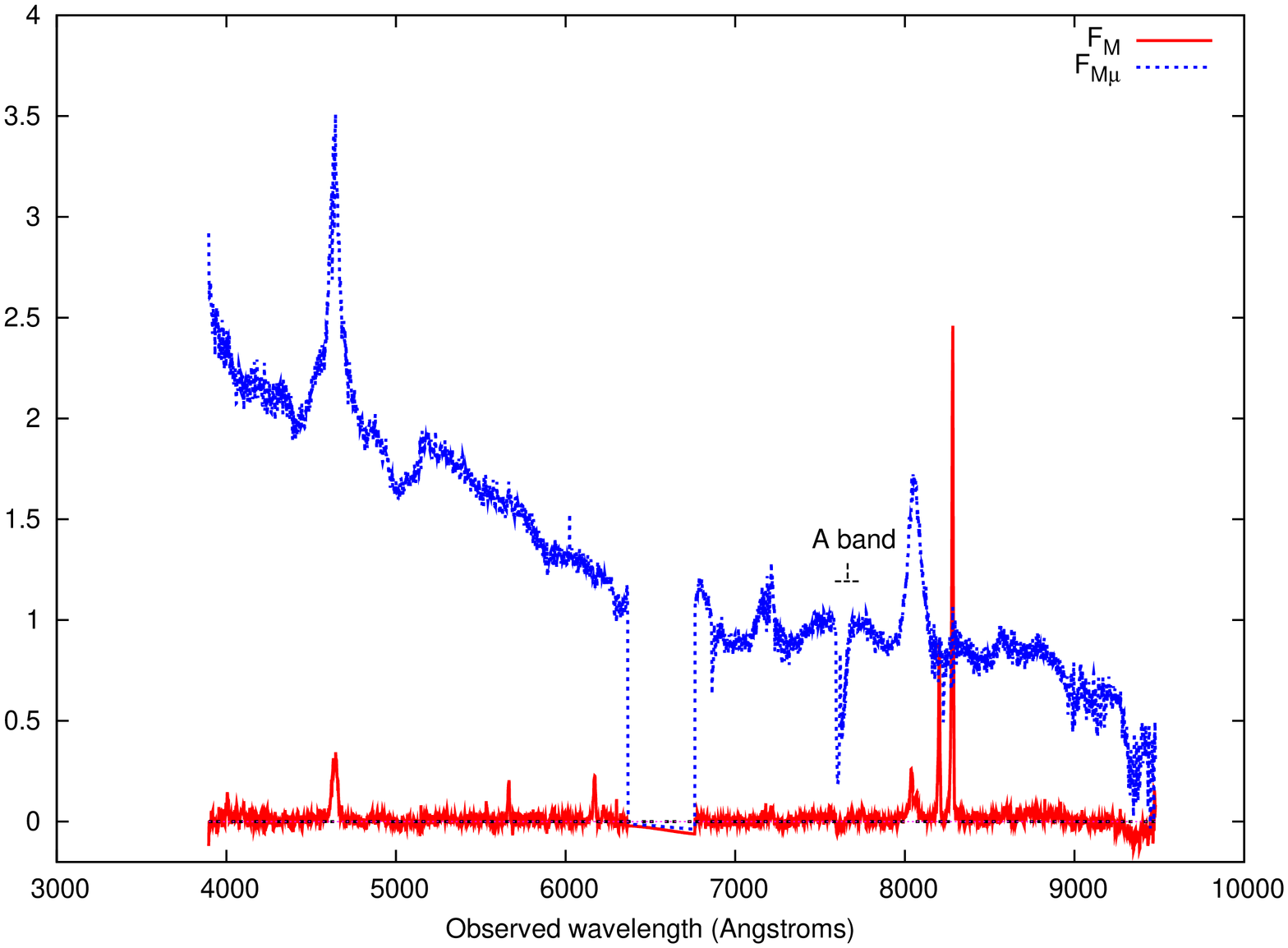}
\caption{Fraction of the spectrum affected ($F_{M\mu}$, dotted) and unaffected ($F_M$, solid) by microlensing (arbitrary units) in the B-C spectrum pair, after removing the observed host contamination (see Eq.~(\ref{eq:decomp2})). The A-band telluric absorption is also identified.}
\label{fig:fmfmuBC}
\end{center}
\end{figure*}

In the case of the B-C pair of spectra, before applying our extraction
method, we must first discuss the {\it chromatic} trend observed in
the ratio spectrum (see Fig. \ref{fig:spec_ratio}). This trend
probably comes from image C since the A/B ratio is not affected. It
has four possible origins: i- chromatic microlensing; ii-
differential extinction along the line-of-sight; iii- host
contamination; iv- contamination by the lensing galaxy.

We exclude significant contamination by the lensing galaxy that is
about 100 times fainter than image C at the lensed images position
(Paper II). Since none of the other phenomena taken separately is
capable of accounting for the observations, we must search for the
minimum combination of hypotheses. Indeed, extinction alone is unable
to explain the presence of \MgII~in the ratio spectrum and is expected
to modify more strongly the spectrum slope in the blue than in the
red. Chromatic microlensing is also expected to mainly affect the blue
part of the spectrum; finally, the host galaxy is expected to maximize
its contamination in the reddest part of the spectrum of the faintest
image, as observed, but it is not supposed to produce a differential
effect on \MgII.

Since the host is indeed observed to be quite bright in the red and in
the NIR (see Papers I \& II), we decided to explore the
combined effect of the host contamination together with a {\em
  wavelength independent} microlensing, as modelled in
Sect.~\ref{sec:decomp}.

The host contamination probably seen in the spectrum ratio necessarily
comes from a constant, {\em resolved} contribution (the possibly {\em
  unresolved} host contribution undergoes the same macro-amplification
as the QSO and is spatially too large to be microlensed, so that it
cannot affect the slope of the spectrum ratio). The contamination by
such a {\em resolved} contribution is {\em additive} and, thus, it
must first be removed before applying Eqs. (\ref{eq:decomp2}). Since,
conversely, the dominant part of the {\em resolved} flux may be
associated with the host and the dominant part of the {\em unresolved}
flux may be associated with the QSO continuum, we can write with a
good approximation:

\begin{equation}
\begin{array}{l}
F_B = F_{\rm QSO} + F_{\rm Host}\\
F_C = R  F_{\rm QSO} + F_{\rm Host}\,,\\
\end{array}
\label{eq:host}
\end{equation}

\noindent
where $R$ is the intrinsic C/B ratio due to macro- and
micro-amplification. $F_{\rm Host}$ is obtained after smoothing out
residual features in the solution of system \ref{eq:host}.

Now, the exact choice of the $R$ value is somewhat arbitrary. If the
host contamination in the blue is assumed to be negligible, the value
$R=0.35$ is chosen, as observed in the blue part of the spectrum ratio
(see Fig. \ref{fig:spec_ratio}).  A break is seen in the
resulting host restframe ``spectrum'' exactly at the expected position
of the Balmer discontinuity, giving a good hint that we indeed deal
with the host galaxy (unfortunately the break falls in the gap between
the two wavelength ranges covered by our spectra). As a refinement,
adopting $R=0.33$ instead of $R=0.35$ adds a slight host contamination
in the blue, so that the final contaminating ``spectrum'' is
reminiscent of the spectrum of a Sb galaxy (see
Fig. \ref{fig:host}). In the first (resp. second) case, the host flux
at 5000\,\AA~ restframe corresponds to $\sim 50$\% (resp. $\sim75$\%)
of the QSO flux observed in image C at the same wavelength. However,
in both cases, the C/B spectrum is flattened and no significant
difference is observed in the subsequent analysis. Note also that
correcting {\em a posteriori} for the host contamination in the A-B
pair does not significantly affect the results derived in
Sect. \ref{sec:ABpair}.

\begin{table}[tb]
\caption{Summary of the macro-amplification ratios $M$, micro-amplification $\mu$ and continuum ratio $R$ values found in the exploratory study.}
\begin{center}
\begin{tabular}{lcc}
\hline \hline
      & A/B & C/B \\
\hline
$M$ & 2.1 & 1.4 \\
$\mu$& 0.30 & 0.25 \\
$R$ & 0.63 & 0.33 \\
\hline
\end{tabular}
\label{tab:sum_explore}
\end{center}
\end{table}

Once the B and C spectra are decontaminated (using $R=0.33$), the
spectral decomposition is performed in the same way as for the A-B
pair. Associating image C (resp. B) with number 1 (resp. 2) in
Eqs. (\ref{eq:decomp2}), we find $M=M_C/M_B=1.4$ and
$\mu=\mu_C/\mu_B=0.25$; the extracted spectra are shown in
Fig. \ref{fig:fmfmuBC}. By construction, the unaffected spectrum is
flat. The BEL appears here nearly totally affected by microlensing
(except parts of the core of \MgII~and \hbeta~BELs). This would
confirm that the BLR is very compact. However, as we shall discuss in
Sect.~\ref{sec:discussion}, this decomposition has to be compared with
the one of the A-B pair and interpreted in terms of the micro-lensing
of several lensed images.
 
Before entering into the possible interpretations, we would
like to further confirm the previous results by deriving the different
flux ratios in a more quantitative way, based on a systematic fitting
of the different spectral components (i.e. host, QSO-continuum, QSO
emission lines, ...).

\section{{\em MCD}: Multi Component Decomposition analysis}
\label{sec:multicomp}

Examination of our extracted spectra reveals significant emission
associated with \FeII. An important fraction of the flux of the host
galaxy is also mixed with the flux from the central AGN
(Sect.~\ref{sec:BCpair}). In order to quantitatively disentangle the
different emission components (which might be differently affected by
microlensing), we used a Multi-Component Decomposition ({\it MCD})
approach (e.g. Wills, Netzer and Wills~\cite{WIL85}; Dietrich et
al.~\cite{DIE03}). This method assumes our rest frame spectra to be a
superposition of: (1) a power law continuum ($F_\nu \sim \nu^\alpha$),
(2) a Balmer emission continuum, (3) a pseudo-continuum due to the
merging of \FeII~emission blends, (4) a galactic template for the
AGN's host galaxy, and (5) an emission spectrum due to the other
individual broad emission lines. The fitting of the spectra is a
classical least square minimization using a Levenberg-Marquard based
algorithm adapted from the Numerical recipes routine (Vetterling et
al.~\cite{VET93}). We give hereafter the technical details associated
with this method. We use it to decompose the observed spectra of A-B-C
(Sect.~\ref{subsec:results}) and discuss the errors in
Sect.~\ref{subsec:caveat}.

\subsection{Overview of the method} 

We performed the fit iteratively. First we fitted components (1), (2)
and (4) in three rest-frame windows nearly free of \FeII~emission,
namely \lll 2600-2700\,\AA, 2950-3100\,\AA~and
5400-5500\,\AA~(Natali et al. ~\cite{NAT98}). Secondly, we imposed
a null weight to the emission line regions and fitted components (1)
to (4) using the previous results as initial conditions. Finally we
included the emission lines in the fit.

\subsubsection{The quasar continuum}

We modelled the quasar emission as: 

\begin{equation}
F_\lambda=F_0 \times \frac{\lambda}{\lambda_0}^{-(2+\alpha)}+F^{\rm Bac}_\lambda
\end{equation}  

\noindent where $\lambda_0=5000$\,\AA, $\alpha$ is the canonical power law
index and $F^{\rm Bac}_\lambda$ is the Balmer continuum intensity. To estimate
the Balmer continuum from our fits, we have used the empirical
distribution by Grandi (\cite{GRA82}) which assumes gas clouds of
uniform temperature $T_e $ that are partially optically thick. In
this case the Balmer continuum spectrum ($\lambda \leq 3646$\,\AA) can
be described by:

\begin{equation}
\label{equ:cont}
F^{\rm Bac}_{\lambda}=F^{\rm BE} B_{\lambda}(T_e)(1-e^{-\tau_\lambda}),~~~\lambda \leq \lambda_{\rm BE}
\end{equation}

\noindent with $B_{\lambda}(T_e)$ as the Planck function at the
electron temperature $T_e$, $\tau_\lambda$ as the optical depth at
$\lambda$, and $F^{\rm BE}$ as a normalized estimate for the Balmer
continuum flux density at the Balmer edge $\lambda_{\rm
  BE}=3646$\,\AA. At wavelengths $\lambda > \lambda_{\rm BE}$ higher
order Balmer lines are merging into a pseudo continuum, yielding a
small rise to the Balmer edge. We did not attempt to model this region
and kept $F^{\rm Bac}_{\lambda}=0$ at $4150$\,\AA\,$> \lambda >
\lambda_{\rm BE}$. In order to ease the minimization, we fixed the
electron temperature and the optical depth $\tau_\lambda$ to realistic
values (i.e. $\tau_\lambda =2.0$ and $T_e = 15000$\,K; Wills et
al.~\cite{WIL85}, Dietrich et al.~\cite{DIE03}). The region
3080-3540\,\AA~is imperfectly modelled using
Eq.~(\ref{equ:cont}). We observe a significant excess of emission
w.r.t. the model. This one is likely explained by blended
\FeII~emission for which there is presently no available template and
which are poorly reproduced by photo-ionization models
(e.g. V\'eron-Cetty et al. ~\cite{VER06}). We thus excluded this
region from the fitting procedure.

\subsubsection{The FeII emission}
\label{subsubsec:FeII}

Following many authors (e.g. Wills et al.~\cite{WIL85}, Laor et
al.~\cite{LAO97}, Dietrich et al.~\cite{DIE03}, Tsuzuki et
al.~\cite{TSU06}), we used template spectra of \FeII~emission instead
of theoretical emission models. Namely, we fitted the \FeII~emission
in our spectra by using an UV \FeII~template (Vestergaard \&
Wilkes~\cite{VES01}) and an optical \FeII~template (V\'eron-Cetty et
al.~\cite{VER04}). We also tested the optical template of Boroson and
Green (\cite{BOR92}) but this did not modify significantly our
results. We prefered the optical template of V\'eron-Cetty et
al. especially due to its larger spectral coverage. Since UV photons
are generally absorbed and re-emitted at optical wavelengths (for high
optical depth), optical and UV \FeII~emission are often
anti-correlated (Joly M., private communication). For this reason, we
have not fixed the intensity ratio between the UV and optical \FeII~
templates. The templates have been broadened in the $\log$ space (in
order to keep the velocity width constant over the whole wavelength
domain) using the $FWHM$ of the \hbeta~emission line. The broadened
templates are shown in Fig.~\ref{fig:spectra}.

For technical reasons, in the \FeII$_{\rm UV}$ template of Vestergaard
\& Wilkes, the \FeII~level beneath the central part of the \MgII~
emission (i.e. $2787 < \lambda < 2802$\,\AA) is 0. We have thus
constructed a modified \FeII~template for which we consider a non zero
\FeII~level between $2752 < \lambda < 2832$\,\AA. We estimate this
level by interpolating in the template the \FeII~level between the two
closest maxima enclosing the zero \FeII~hole. These maxima are found
at 2752 and 2832\,\AA. Both templates have been used during our
decomposition. The effect of this modified template on the results is
discussed in Sect.~\ref{subsec:emission}.

\subsubsection{Host galaxy}

We used a Sb galactic template (Kinney \cite{KIN93}) to model the host
galaxy. The latter choice is motivated by the detailed study of the
host performed in Paper II. It is also supported by the empirical host
spectrum we have retrieved in Sect.~\ref{sec:explore} and which has a
continuum emission compatible with a Sb type galaxy
(Fig.~\ref{fig:host}).

\subsubsection{Emission lines} 

During the last iteration of our fitting procedure, the emission lines
were fitted simultaneously with the other components as a sum of
Gaussian profiles. A minimum of two profiles was considered for the
main emission lines. More details are given in
Sect.~\ref{subsec:emission}.

\subsection{Application to A-B-C spectra}
\label{subsec:results}

We present in this section the results of the {\it {MCD}} method
applied to the spectra of images A, B and C in order to
estimate the flux ratios in the different emission
regions. Unfortunately, we found that there is a degeneracy between
three quantities: the host galaxy relative flux, the power law index
$\alpha$ and the Balmer continuum level $F^{\rm Bac}_{\lambda}$. Due
to the large range of acceptable values for $\alpha$ implied by the
degeneracy, we could not disentangle the host galaxy emission from the
QSO emission and we could not assess the presence (absence) of chromatic
variation of the quasar continuum (due either to microlensing or
differential extinction). In the following, we thus decide to make a
prior on $\alpha$ in order to break the degeneracy. This has no major
effects on our conclusions as explained in Sect.~\ref{subsec:caveat}.

Motivated by the results of Sect.~\ref{sec:explore}, we imposed
$\alpha$ to be identical for each image. We choose $\alpha=-0.125$
which is at the middle of the overlapping range of acceptable values
for $\alpha$ in A, B, C, when no prior is imposed. This value is also
similar to the best value found for image B. This value is in good
agreement with the mean power law index $\alpha=-0.33\pm0.59$ found by
Natali et al. (\cite{NAT98}) in the range 1200-5500\,\AA\, or with the
power law index of the composite SDSS quasar spectra measured in the
range 1300-5000\,\AA\, (i.e. $\alpha=-0.44$; Vanden Berk et
al. ~\cite{VAN01}). Results are given in Table~\ref{tab:multicomp}.

\subsubsection{The continuum}
\label{subsec:continuum}

\begin{table*}[]
  \caption{Multi-component best fit of the spectra when $\alpha$ is forced to be identical in A, B, C. Col.~1 is the image ID, Col.~2 is the power law index $\alpha$ of the quasar continuum, Col.~3 gives the flux level of the power law continuum at $\lambda = 5000$\,\AA~(rest-frame), Col.~4 gives the flux level of the Balmer continuum, Col.~5 gives the relative flux of the host galaxy at 5000\,\AA~normalized by $F^{\rm PWL}_0$ (in B), Col.~6 is the level of the optical template of \FeII, Col.~7 is the level of the UV template of \FeII. All the quantities (except $\alpha$) are normalized to image B.} 
\label{tab:multicomp}
\centering
\begin{tabular}{ccccccc}
\hline \hline
image & $\alpha$ & $F^{\rm PWL}_0$ & $F^{\rm Bac}$ & $F_{{\rm host},0}$ & \FeII$_{\rm opt}$ & \FeII$_{\rm UV}$ \\
\hline

A &	-0.125&	0.55$\pm$	0.02&	0.94$\pm$	0.09&	0.58$\pm$	0.02&		1.91$\pm$	0.03&	2.29$\pm$	0.08\\
B &	-0.125&	1.00$\pm$	0.01&	1.00$\pm$	0.09&	0.77$\pm$	0.02&		1.00$\pm$	0.03&	1.00$\pm$	0.08\\
C &	-0.125&	0.31$\pm$	0.01&	0.41$\pm$	0.09&	0.54$\pm$	0.01&		0.43$\pm$	0.03&	0.82$\pm$	0.06\\


\hline


\end{tabular}
\end{table*}

The flux ratios between the power law continua are directly obtained
from the MCD results (Table~\ref{tab:multicomp}) and are reported in
Table \ref{tab:sum_mcd}. We find $F_{A}/F_{B} = 0.55\pm0.02$ and
$F_{C}/F_{B} = 0.31\pm0.01$, in good agreement with the rough
measurements performed on the direct spectral ratios ($F_{A}/F_{B} =
0.63$ and $F_{C}/F_{B} = 0.33$, see Table
\ref{tab:sum_explore}). Since the estimate of the host galaxy
contribution to the observed spectra is very different with both
methods (i.e. 180\% of the QSO flux in C from the {\it MCD} method and
75\% from the direct ratio), we are confident that the error on the
host galaxy level does not dramatically bias our continua flux
ratios{\footnote{Note that it neither affects the $F\mu$ decomposition
    of the spectra. Namely, a similar $F\mu$ decomposition is obtained
    with these two different host estimates.}.

\subsubsection{The emission lines}
\label{subsec:emission}

\begin{table*}
  \caption{Multi-component fit of the main emission lines normalized to the total flux in \hbeta~for image A. Note that \halpha~normalization is arbitrary due to the wrong flux scaling of the NIR spectra of J1131 w.r.t. the optical spectra.}
  \centering
  \begin{tabular}{ll|cccccc}
    \hline
&ID & \ll$_c$ &  $FWHM$ & $F_{\rm A}$ & $F_{\rm B}$ & $F_{\rm C}$ & $z$ \\
\hline
\MgIIl	&BC1&4636.4	&2050	&	0.281$\pm$0.006	&0.077$\pm$0.006	&0.081$\pm$0.004	&	0.657\\
	&BC2&4638.4	&4320	&	0.867$\pm$0.009	&0.557$\pm$0.008	&0.251$\pm$0.006	&	0.658\\
	&VBC&4629.4	&21830  &	0.728$\pm$0.013	&0.646$\pm$0.009	&0.215$\pm$0.006	&	0.655\\
        &TOT&	        &	&	1.842$\pm$0.017	&1.253$\pm$0.014	&0.534$\pm$0.009	&            \\
        & no VBC&	&	&	1.148$\pm$0.010	&0.634$\pm$0.010	&0.332$\pm$0.007	&            \\
\hbetal	&NC1&8039.6	&460	&	0.040$\pm$0.001	&0.012$\pm$0.001	&0.016$\pm$0.001	&	0.654\\
	&BC1&8054.2	&2000	&	0.345$\pm$0.005	&0.134$\pm$0.005	&0.091$\pm$0.004	&	0.657\\
	&BC2&8055.4	&4310	&	0.615$\pm$0.006	&0.477$\pm$0.006	&0.159$\pm$0.005	&	0.657\\
        &TOT&	        &	&	1.000$\pm$0.008	&0.623$\pm$0.008	&0.266$\pm$0.006	&            \\
        &BC &	        &	&	0.960$\pm$0.008	&0.611$\pm$0.008	&0.250$\pm$0.006	&            \\
\halphal&NC1&10850.8    &550	&	0.496$\pm$0.015	&0.242$\pm$0.015	&0.276$\pm$0.015	&	0.653\\
	&NC2&10831.7    &550	&	0.591$\pm$0.013	&0.244$\pm$0.013	&0.293$\pm$0.013	&	0.654\\
	&NC3&10884.8    &550	&	0.199$\pm$0.018	&0.158$\pm$0.018	&0.337$\pm$0.018	&	0.653\\
	&BC1&10874.5    &2070   &	3.595$\pm$0.063	&1.686$\pm$0.063	&1.122$\pm$0.063	&	0.657\\
	&BC2&10872.5    &4300   &	5.480$\pm$0.078	&4.227$\pm$0.079	&1.533$\pm$0.080	&	0.657\\
        &TOT& 	        &	&	10.361$\pm$0.104&6.557$\pm$0.104	&3.562$\pm$0.105	&            \\
        &BC&    	&	&	9.075$\pm$0.101	&5.913$\pm$0.101	&2.655$\pm$0.102	&            \\
& & & & & & &\\
\OIIId	&NC1&8202.1	&240	&	0.052$\pm$0.001	&0.031$\pm$0.001	&0.038$\pm$0.001	&	0.654\\
	&NC2&8198.8	&660	&	0.073$\pm$0.002	&0.034$\pm$0.002	&0.049$\pm$0.002	&	0.653\\
	&NC3&8280.7	&310	&	0.214$\pm$0.002	&0.110$\pm$0.001	&0.150$\pm$0.002	&	0.654\\
	&NC4&8275.6	&800	&	0.172$\pm$0.003	&0.084$\pm$0.002	&0.108$\pm$0.002	&	0.653\\
        &TOT&   	&	&       0.511$\pm$0.004&0.259$\pm$0.003	        &0.345$\pm$0.003	&	     \\
\NeV\,\ll3425&NC1&5664.3&520	&	0.033$\pm$0.002	&0.017$\pm$0.002	&0.019$\pm$0.001	&	0.653\\
\OIIl   &NC1&6171.6	&620	&	0.023$\pm$0.001	&0.013$\pm$0.001	&0.023$\pm$0.001	& 	0.656\\
\hline                                                                            
  \end{tabular}                                                             
  \label{tab:multicompfitfix}
\vfill
\end{table*}

\begin{figure*}[t]
\centering
 \includegraphics[angle=-90, width=1.0\columnwidth]{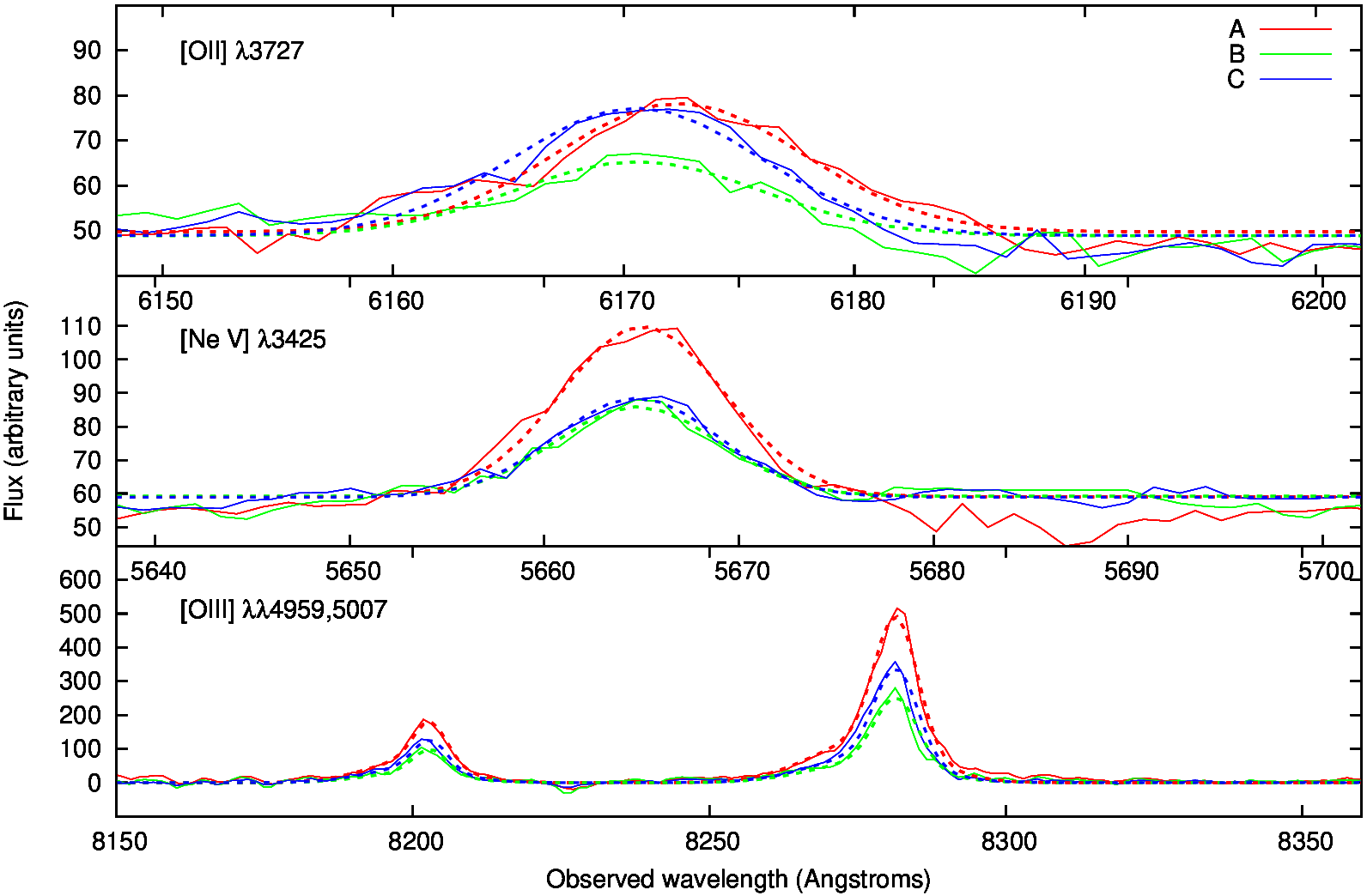}
  \includegraphics[angle=-90, width=1.0\columnwidth]{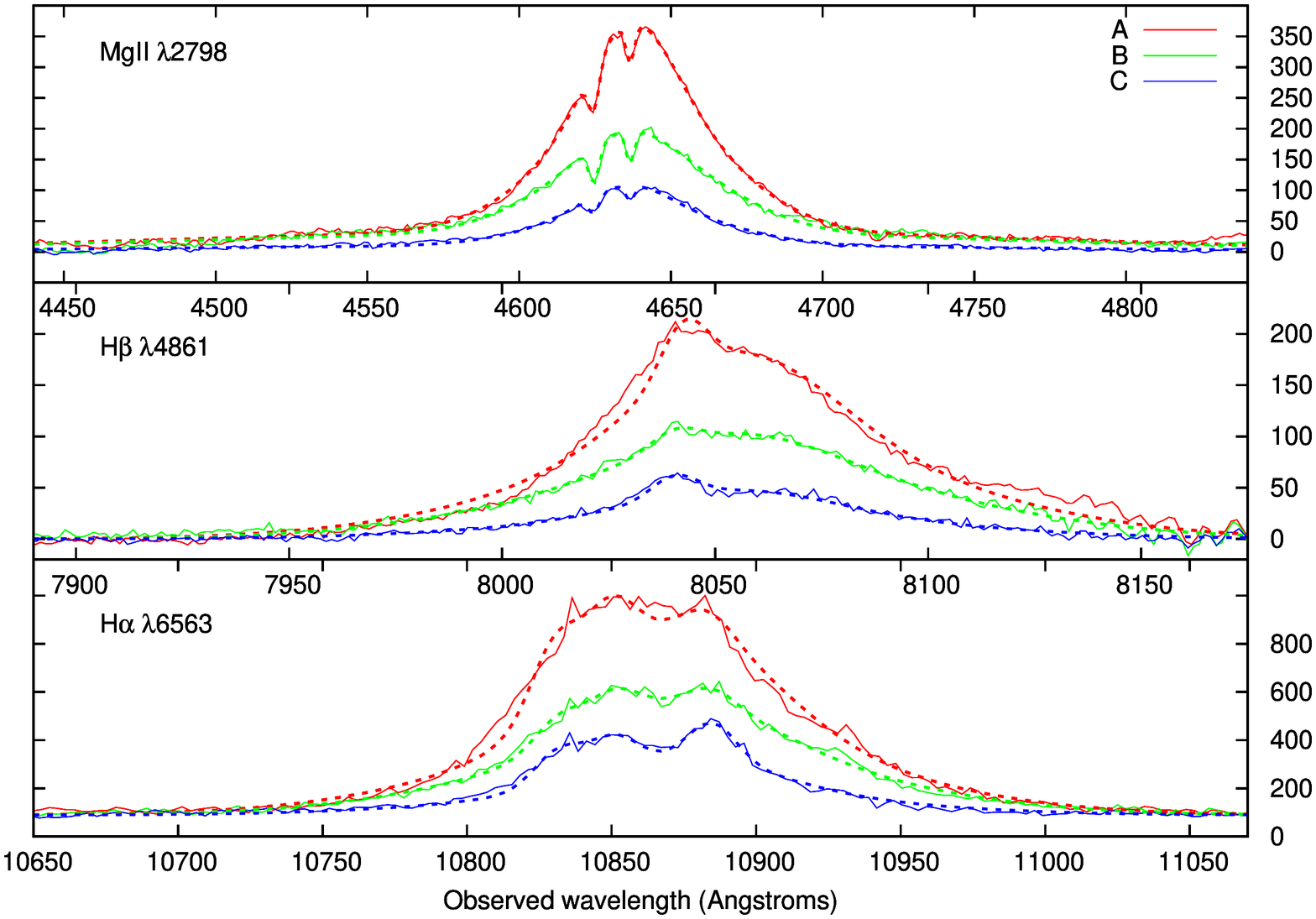}
  \caption{ Comparison of the main NEL ({\it left panel}) and BEL ({\it right}) for
    A (solid red), B (solid green) and C (solid blue) with their
    multi-component fitted model (superimposed with dotted lines).
    \hbeta, \MgII~and \OIIId~are continuum subtracted profiles. For
    \halpha, \NeV~and \OII~, we have normalized the apparent pseudo
    continuum to the one of image C. }
  \label{fig:emissionmulti}
\end{figure*}

Faint NELs were fitted with a single Gaussian{\footnote{Due to the low
    contrast between \NeIV~$\lambda$2419 and \hgamma+\OIII \lll 4340,
    4363 emissions w.r.t. their underlying continuum, the fit of those
    lines is inaccurate and not presented.}} profile while the bright
\OIIId~were fitted with 2 components. The latter decomposition enables
us to take into account the line asymmetry of the \OIIId~ emission
lines. Since the $FWHM$ of the narrowest component is close to the
spectral resolution, one should not over-interpret this
decomposition. In the following, we will not use the flux measured in
the individual components but rather the total flux in the line.

Regarding the BELs, as it is a common practice, we decomposed each
line into two broad Gaussian profiles (e.g. V\'eron-Cetty \&
V\'eron~\cite{VER00}, Sulentic et al.~\cite{SUL00}, Romano et
al.~\cite{ROM96}) and we added a narrow component when necessary.  We
enforced the $FWHM$ and $\lambda_c$ of the various sub-profiles to be
identical for each lensed image. Motivated by the claimed similarity
between the \MgII~and Balmer line profiles (e.g. Mc Lure \&
Jarvis~\cite{MCL02}), we enforced the velocity width of the 2 broad
components (i.e. $FWHM$ $\sim 2000$ and $4300$\,km s$^{-1}$) to be
identical -within 50\,km s$^{-1}$- for each broad line. However, like
e.g. Corbett et al (\cite{COR03}), we found necessary to add a very
broad component ($FWHM$$\sim 22\,000$\,km s$^{-1}$) to the \MgII~profile. We
did not find any evidence for such a component in the Balmer line
profile. We supposed that the different decomposition found for \MgII~
could be due to the hole of \FeII~beneath \MgII~in the \FeII$_{\rm
  UV}$ template. However, the use of a modified \FeII~template in
which we ``filled'' the \FeII~hole (Sect.~\ref{subsubsec:FeII}) does
not prevent the presence of a very broad component in the \MgII~
profile. The fit of \MgII~was even found to be worse with this
modified template and especially, the peak of the \MgII~emission could
not be fitted correctly anymore. Finally, we modelled the faint
\MgII~narrow absorption line doublet seen in the \MgII~broad emission
with 2 Gaussian profiles separated by 7.2\,\AA~rest frame
(Sect.~\ref{subsec:absorption}).

For images B and C, our best fitted profile of the broad emission
lines was indistinguishable (in terms of $\Delta \chi^2$) from the
profile obtained when no prior was imposed on the decomposition. For
image A, when no prior was imposed on the fit, we found that the best
decomposition of the Balmer lines was made of two broad profiles
separated by typically 1000\,km s$^{-1}$ suggesting some {\it asymmetry} in
the line profile.

We report in Table~\ref{tab:multicompfitfix} the results of the fits
performed with the {\it MCD} method.  Figure~\ref{fig:emissionmulti}
shows the continuum subtracted emission lines in each lensed image,
together with their modelled profile.

The flux ratios calculated using the results of
Table~\ref{tab:multicompfitfix} are similar for \halpha, \hbeta~and
\MgII~BELs. The flux ratios calculated for the narrow \OIIId~and
\NeVb~emission lines agree within the error bars but are different
from those obtained in the BELs. This independently confirms our
previous finding that both $F_A/F_B$ and $F_C/F_B$ have different
values for the BELs and for the NELs. Note that due to the excellent
agreement between the flux ratios measured in \OIII$\lambda$4959
(NC1+NC2) and \OIII$\lambda$5007 (NC3+NC4), we do not distinguish
between these two lines in the following and rather consider the total
flux \OIII$_{\rm tot}$ measured in the \OIII~doublet.

The most precise flux ratio estimates are obtained for the broad
component BC of \hbeta\ and for \OIII$_{\rm tot}$. They represent
characteristic values of the BEL and the NEL respectively and are
reported in Table~\ref{tab:sum_mcd}. We note that the \OIII~flux ratio
values ($ F_A/F_B($\OIII$)=1.97$ and $F_C/F_B($\OIII$)=1.33$) found
here are in good agreement with the macro-amplification ratios derived
in Sect.~\ref{sec:explore} to match the \OIII~emission in each
spectrum (see Table~\ref{tab:sum_explore}). 

Interestingly, for each BEL, the flux ratios found for the broadest
components (e.g. BC2 or VBC) are different from those of the narrowest
broad component (BC1) except for $F_A/F_C$.

The \FeII$_{\rm opt}$ flux ratio (Table~\ref{tab:multicomp}) is
similar to the \hbeta~ratio for the image pair B-C and to the
\OIII~ratio for the image pair A-B, as can be easily seen in Table
\ref{tab:sum_mcd}.  There is an indication for a similar behaviour in
\FeII$_{\rm UV}$ but the flux ratios deviate more significantly from
those obtained in \hbeta~ and \OIII.

Finally, a systemic redshift $z_{\rm {systemic}}=0.657\pm0.001$ has
been determined from the \MgII~emission. The measured narrow emission
lines are all blue-shifted w.r.t. the systemic redshift
($z_{NEL}=0.654\pm0.001$) except \OII~emission
($z_{[OII]}=0.656\pm0.001$). We also notice that the red wing of the
emission line profile of \OII~ is significantly more pronounced in
image A (Fig.~\ref{fig:emissionmulti}), probably because of some
underlying \FeII~emission (which may deviate from the template we
used) and/or to emission associated with star formation located in a
direction orthogonal to the slit.

In summary, the {\it MCD} analysis performed in this section confirms
the main results unveiled in Sect.~\ref{sec:explore}. First, we
measure significantly different flux ratios for the continuum
emission, the BELs and the NELs. Secondly, we confirm differential
magnification of the broadest component of the BELs compared to the
core of the emission lines as shown in
e.g. Fig.~\ref{fig:fmfmuAB}. Finally, we find that the flux ratio
$F_C/F_B$ (resp. $F_A/F_B$) in \FeII~is similar to the one of the BEL
(resp. NEL).

\begin{table}[h]
\caption{Summary of the flux ratio values measured with the {\it MCD} method for several important spectral features.}
\begin{center}
\begin{tabular}{lcc}
\hline \hline
      & A/B & C/B \\
\hline
\OIII & 1.97$\pm$0.03 & 1.33$\pm$0.02\\
\hbeta& 1.57$\pm$0.06 &  0.41$\pm$0.01\\
\FeII$_{\rm opt}$ & 1.91$\pm$0.08 & 0.43$\pm$0.04 \\
Continuum & 0.55$\pm$0.02 & 0.31$\pm$0.01 \\
\hline
\end{tabular}
\label{tab:sum_mcd}
\end{center}
\end{table}

\subsection{Error analysis}
\label{subsec:caveat}

The {\it MCD} method provides us with 1$\sigma$ formal errors
calculated from the covariance matrix in the Levenberg-Marquard
minimization routine (Vettering et al.~\cite{VET93}). The main
limitation of the {\it MCD} technique comes from a degeneracy between
the power law index $\alpha$ (and the level $F^{\rm PWL}_0$), the
Balmer continuum emission ($F^{\rm Bac}$) and the galactic template
contribution. This means that these quantities cannot be retrieved
unambiguously. Before fixing the same value for $\alpha$ in each
image, we have estimated the range of acceptable values for $\alpha$
and $F^{\rm PWL}_0$ by fixing the host galaxy level to two extreme
values beyond which the fit becomes unacceptable. For that range of
acceptable continuum spectral decompositions, we found that the fit of
the \MgII, \hbeta, \OIIId~and \FeII$_{\rm opt}$ remained unchanged. On
the other hand, the \FeII$_{\rm UV}$ intensity was retrieved with only
20\% accuracy. Consequently, imposing the same value for $\alpha$ in
each lensed image is safe and does not induce systematic error on the
fit of the emission lines (including \FeII).

However, we want to be more cautious regarding the results of the \NeV
\ll3425, \OIIl~emission lines. Indeed, we have no satisfying model for
the pseudo continuum under these lines. Therefore, the Gaussian
fitting of these lines is performed independently using a flat
pseudo-continuum under the line. Because of the likely presence of
\FeII~in that pseudo-continuum, our procedure may introduce a
systematic bias on the fitted line intensity.

\section{Towards a microlensing scenario} 
\label{sec:discussion}

\begin{table}[]
\begin{minipage}[t]{\columnwidth}
\caption{Micro-amplification ratio $\mu_i/\mu_j$ between images $i$ and $j$ calculated by normalizing the flux ratio measured for a given emission region by the flux ratio measured in \OIIId~(Sect.~\ref{sec:discussion}). For completeness, we have also added the micro-amplification ratios computed from optical and X-ray imaging at different epochs (mm/yy). Note that $FWHM$$(BC1)\sim 2050$\,km s$^{-1}$; $FWHM$$(BC2)\sim 4300$\,km s$^{-1}$; $FWHM$$(VBC)\sim 22\,000$\,km s$^{-1}$.}
\label{tab:fluxratios}
\centering
\renewcommand{\footnoterule}{}  
\begin{tabular}{lccc}
\hline \hline
ID                      & $\mu_A/\mu_B$	& $\mu_A/\mu_C$	& $\mu_C/\mu_B$	\\
\hline
\OIIId$_{\rm tot}$	&1.00$\pm$	0.02	&1.00$\pm$	0.03	&1.00$\pm$	0.02\\
\NeV$\lambda$3425	&0.97$\pm$	0.10	&1.16$\pm$	0.13	&0.83$\pm$	0.11\\
\OII$\lambda$3727	&0.88$\pm$	0.11	&0.68$\pm$	0.08	&1.29$\pm$	0.12\\
\MgIIl \hglue 1mm	BC1	        &1.85$\pm$	0.14	&2.33$\pm$	0.21	&0.79$\pm$	0.09\\
\hglue 16mm	BC2	        &0.79$\pm$	0.02	&2.34$\pm$	0.08	&0.34$\pm$	0.03\\
\hglue 16mm	VBC	        &0.57$\pm$	0.02	&2.29$\pm$	0.09	&0.25$\pm$	0.04\\
\hglue 16mm     TOT	&0.75$\pm$	0.02	&2.33$\pm$	0.07	&0.32$\pm$	0.03\\
\hglue 16mm     no VBC	&0.92$\pm$	0.02	&2.34$\pm$	0.08	&0.39$\pm$	0.03\\
\hbetal	\hglue 2mm NC1	        &1.72$\pm$	0.21	&1.66$\pm$	0.23	&1.03$\pm$	0.14\\
\hglue 15mm	BC1	        &1.31$\pm$	0.05	&2.56$\pm$	0.15	&0.51$\pm$	0.06\\
\hglue 15mm	BC2	        &0.65$\pm$	0.01	&2.61$\pm$	0.10	&0.25$\pm$	0.04\\
\hglue 15mm     TOT	&0.82$\pm$	0.02	&2.54$\pm$	0.08	&0.32$\pm$	0.03\\
\hglue 15mm     {\bf BC}	&{\bf 0.80$\pm$	0.02}	&{\bf 2.59$\pm$	0.09}	&{\bf 0.31$\pm$	0.03}\\
\FeII$_{\rm	opt}$	&0.97$\pm$	0.04	&3.04$\pm$	0.28	&0.32$\pm$	0.10\\
\FeII$_{\rm	UV}$	&1.16$\pm$	0.12	&1.89$\pm$	0.18	&0.61$\pm$	0.12\\
{\bf Continuum}		&{\bf 0.28$\pm$	0.01}	&{\bf 1.23$\pm$	0.07}	&{\bf 0.23$\pm$	0.04}\\
$R$ band 04/03	        &0.31	&1.24	&0.26\\
$R$ band 04/04 {\footnote {Paper I}}	        &0.46	&1.69&0.27\\
$R$ band 04/05 {\footnote {Morgan et al. (\cite{MOR06})}}	        &0.58&1.43	&0.41	\\
X Ray 04/04 {\footnote {Pooley et al. (\cite{POO06})}}	        &0.09&0.45	&0.20	\\
\hline

\end{tabular}
\end{minipage}
\end{table}

In this section we look whether the difference in flux ratios observed
in the NELs, the BELs and continuum (Sect.~\ref{sec:explore} \&
\ref{subsec:results}) can be understood by means of micro-lensing
occurring in one or several lensed images. We first identify a possible
scenario following a quantitative approach and then discuss the implications.

\subsection{A possible microlensing scenario}

Because the narrow line emission takes place in a region larger than
the BLR and the continuum, it is likely to be much less affected by
micro-lensing. Therefore, as already mentioned in
Sect. \ref{sec:explore}, we can reasonably assume that the
macro-amplification ratios are close to the flux ratios{\footnote
    {The flux ratios measured in \OIIId ($F_A/F_B=1.97$,
    $F_C/F_B=1.33$) slightly differ from those obtained with the
    SIE+$\gamma$ fiducial model used in Paper I ($F_A/F_B=1.65$,
    $F_C/F_B=0.9$) indicating some inadequation between the model and
    the data (see Sect.~\ref{subsec:macroratio}).}} in \OIIId. So, we
can write:

\begin{equation}
\frac{M_i}{M_j} = \frac{F_i}{F_j}([{\rm {O}}_{\rm III}]),
\end{equation}

\noindent 
where $M_i$ ($M_j$) is the macro-amplification of image $i$~($j$) and
$F_i/F_j$ is the flux ratio in \OIII\ calculated for images $i$ and $j$. On
the other hand, if micro-lensing $\mu_i$ ($\mu_j$) occurs in image
$i$($j$), we can write for a given emission region:

\begin{equation}
\frac{F_i}{F_j}({\rm emission})=\frac{\mu_i}{\mu_j}\frac{M_i}{M_j}.
\end{equation}

\noindent Consequently, if we normalize the flux ratio estimated for a given
emission region by the flux ratio in \OIIId, we infer the
micro-amplification ratio $\mu_i/\mu_j$ for that
region. Table~\ref{tab:fluxratios} displays the micro-amplification
ratio $\mu_i/\mu_j$ for the main emission regions computed using the
results of Table~\ref{tab:multicompfitfix} (we have not reported the
results for \halpha~because they are similar to those obtained for
\hbeta). A first quick look at Table~\ref{tab:fluxratios} shows that,
except for the NELs (namely \OII, \NeV), the $\mu_i/\mu_j$ ratio is
{\it not} compatible with 1 for any value of ($i,j$). This confirms
what we had already revealed in Sect.~\ref{sec:explore}, namely that
micro-lensing is occurring {\it in more than one image}.

We now examine whether the results are compatible with microlensing
occurring in {\it two} images. A priori, there are three possible
scenarios, depending on whether the un-affected image is assumed to be
A, B or C. Thus, each scenario corresponds to $\mu_A=1$, $\mu_B=1$ or
$\mu_C=1$, respectively. From Table~\ref{tab:fluxratios} it is then
easy to derive the respective micro-amplification values
($\mu_B,\mu_C$), ($\mu_A,\mu_C$) and ($\mu_A,\mu_B$), both in the
continuum and in the broad components BC of \hbeta~(see values in bold
face in Table~\ref{tab:fluxratios}). A robust rule to reject a
scenario is the following: the BLR being spatially larger than the
continuum emitting region, it cannot be {\it more} affected by
microlensing than the latter. In other words, the micro-amplification
value must be closer to 1 in the BLR for the two affected images in
the considered scenario. It is straightforward to check that the above
rule is satisfied only in the case when {\it image B is not
  affected}. We find in that case that ($\mu_A,\mu_C$) = (0.28, 0.23)
in the continuum and ($\mu_A,\mu_C$) = (0.80, 0.31) in \hbeta, which
means a {\it de-amplification} of both images A and C. Both the
continuum emitting region {\it and} the BLR are affected. We now
discuss these results.

\subsection{Micro-lensing of the BLR}

As already stated in the introduction (Sect.~\ref{sec:intro}),
microlensing of the BLR is not surprising in the case of J1131. Here
we demonstrate that microlensing has indeed the correct scale to
affect the BLR. The Einstein radius $R_E$ of a star of mass $M$ is
given by:

\begin{equation}  
\label{equ:micro}
R_E= \sqrt{\frac{4GM}{c^2}\frac{D_{ls}D_{os}}{D_{ol}}}=14.3 \sqrt{\frac{M h^{-1}}{M_{\odot}}}\,{\rm lt-days},
\end{equation}

\noindent
where $D_{os}$, $D_{ls}$, $D_{ol}$ are the angular-size distances
between observer and source (os), lens and source (ls) and observer
and lens (ol).

On the other hand, Kaspi et al. (\cite{KAS05}) have deduced a relation
between the size of the BLR and the luminosity
$L_{\lambda}(5100\,\AA)$ of the QSO. An improved relation correcting
from the host galaxy contamination has been derived in Bentz et
al. (\cite{BEN06}). Using the magnitude of image B in the F814W filter
measured with HST (Paper II), and assuming an amplification factor of
10 for that image, we can estimate the likely size of the BLR using
Bentz et al.'s relation. We find that the luminosity $\lambda
L_{\lambda}$(4900\,\AA) = 7.91 10$^{43} h^{-2}$ erg/s which
translates into a likely size for the \hbeta~emission line region of
$\sim$ 40-50 lt-days (using $h=0.7$; 20-40 lt-days using Kaspi et
al.'s relation). Although this can only be a rough estimate due to the
bias on the luminosity introduced by the presence of \OIII~+
\hbeta~emission in the F814W filter and due to the uncertainty on the
macro-lens amplification, we see that a few solar mass micro-lens has
an Einstein radius large enough to affect a significant fraction of
the broad emission line region.

\subsection{Micro-de-amplification of image A}

The micro-de-amplification we find for image A is an independent
confirmation of the results found on the basis of flux ratio temporal
variability in the optical (Paper I). They also fully support X-ray
observations. Indeed, because there is evidence that the QSO X-ray
emitting region is smaller than the continuum emitting region
(e.g. Pooley~\cite{POO06}), the de-amplification is expected to be
stronger in the X-ray than in the optical, at a given epoch. This was
clearly the case in April 2004, with $\mu_{A,X}=0.09$ and
$\mu_{A,R}=0.46$, as reported in Table~\ref{tab:fluxratios} from data
in Pooley et al. (2006) and in Paper I, respectively. Finally, the
parity of image A being negative (saddle point image),
de-amplification can easily be produced by substructure(s) of any mass
(e.g. Keeton~\cite{KEE03}). Given the time variability detected in
Paper I, the microlensing regime is favoured rather than the
milli-lensing one.

\subsection{Micro-de-amplification of image C}

Like for image A, the de-amplification of image C is possibly
supported by the even stronger de-amplification observed in the X-ray
domain in April 2004 (i.e. $\mu_{C,X}=0.20$ and $\mu_{C,R}=0.27$, see
Table~\ref{tab:fluxratios}). Note however that while the X-ray data
alone cannot discriminate between the micro-de-amplification of C and
the micro-amplification of B, our data only support the former
scenario. A sharp drop in the flux of image C has also been observed
between 2004 and 2006 in the X-ray domain (Kochanek et
al.~\cite{KOC06}). Of course, this is not evidence for
de-amplification in 2003 but simply another sign that C can undergo
variable microlensing.

Unlike the parity of image A, the parity of image C is {\it positive}
(minimum image). Consequently, it can only be {\it amplified} by an
{\it isolated} substructure. This is not only the case with an
isothermal clump (Keeton~\cite{KEE03}) but also with a point-mass
perturber in the scenario of micro-lensing (i.e. in the
Chang-Refsdal~\cite{CR84} lens model). This is so because for a
minimum image and $\kappa < 1$, we always have $\kappa + \gamma <1$,
where $\kappa$ and $\gamma$ are the macro model dimensionless surface
mass density and shear values at the position of the considered
image. Adding a second point mass perturber does not change the result
(Grieger et al.~\cite{GRI89}). However, if a non-negligible fraction
of the surface mass density of the lens is made of micro-deflectors,
the probability to {\it de-amplify} a minimum macro-image increases a
lot, as illustrated in Fig. 3 of Schechter and
Wambsganss~(\cite{SCH02}). Their M10 model (macro-amplification of 10
of a minimum image) represents a situation quite similar to the one of
the C image of J1131. The observed de-amplification of image C by a
factor $\sim 4$ would constrain the fractional contribution of
micro-lenses to be larger than 30\% of the total surface mass
density of the lens around image C.

\subsection{Conclusion}

We have shown that the most simple micro-lensing scenario consistent
with the published data on J1131 implies that images A and C are both
de-amplified by a micro-lens. The latter also affects the BELs. We
have presented supporting arguments of this scenario.

\section{Microlensing as a probe of the QSO structure}
\label{sec:probe}

In the previous section, we presented a scenario accounting for the
discordant flux ratios measured in the continuum, the BELs and the
NELs. We have shown that the data are compatible with microlensing
de-amplification of images A and C. Interestingly, both our
exploratory and multi-component analyses (Sects.~\ref{sec:explore} \&
\ref{subsec:results}) show that the BEL is nearly completely
micro-lensed in image C and partially micro-lensed in image A. This
difference of behaviour is likely due to the larger Einstein radius
$R_E$ of the micro-lens affecting C. Indeed, the micro-lensed fraction
of a given emitting region increases with the $R_E$ of the micro-lens
(Eq.~\ref{equ:micro}). In this context, an emission region that is
micro-lensed in image A should also be micro-lensed in image C and
should be very compact, while a larger emitting region might be
micro-lensed only in image C. Finally, even larger regions should not
be micro-lensed either in A or in C. Consequently, we are able to
infer information about the BEL structure by simply identifying and
characterizing the emitting regions microlensed in A \& C.

This scenario further supports our assumption that there is no
chromatic micro-lensing of the optical continuum. Indeed, chromatic
micro-lensing is expected to be stronger for smaller Einstein radii as
well as during caustic crossing events (Wambsganss \&
Paczy\'nski~\cite{WAM91}). Due to the smaller $R_E$ in image A than in
C and to the absence of time flux variations of the B/C ratio between
2002 and 2004 (Paper I) -indicating that there was no caustic crossing
event for image C-, we would expect chromatic micro-lensing of the
optical continuum to be stronger in image A than in C. However, the
absence of a chromatic trend in the A/B spectral ratio
(Fig.~\ref{fig:spec_ratio}) suggests that this effect is
negligible. It is thus meaningful to phenomenologically probe the BLR
structure using the {\it {MCD}} analysis of A-B-C
(Sect.~\ref{subsec:results}) and the results of the $F\mu$
decomposition (Sect.~\ref{subsec:exploreFmu}).  The adopted
micro-lensing scenario provides us with a simpler interpretation of
the $F\mu$ decomposition of the A-B and B-C pairs of spectra. Indeed,
the absence of micro-lensing in image B means that the $F\mu$
decomposition of resp. the A-B and B-C pairs unveils the micro-lensed
regions in resp. A and C.

We discuss in the next two sections the structural information inferred
from the microlensing of the emission line regions
(Sect.~\ref{sec:microBEL}) and of the pseudo-continuum emitting region
(Sect.~\ref{sec:microFeII}).

\subsection{Microlensing of the emission line regions}
\label{sec:microBEL}

A first interesting result relates to the differences existing between the
\MgII~and Balmer emission lines. Indeed, the assumed decomposition of
the Balmer lines with 2 broad components (BC1 with
$FWHM$\,$\sim2000$\,km s$^{-1}$ and BC2 with $FWHM$\,$\sim4300$\,km s$^{-1}$) did
not enable us to reproduce the \MgII~profile. The foot of that line is
clearly larger than the one of the Balmer lines and is well modelled
with a very broad component VBC
($FWHM$\,$\sim22\,000$\,km s$^{-1}$). Additionally, we found with the {\it MCD}
method that VBC is significantly de-amplified in {\it both} A and C
($\mu_A(\rm {VBC})=0.57$; $\mu_C({\rm VBC})=0.25$;
Table~\ref{tab:fluxratios}). This is confirmed by the finding of a
very broad component in $F_{M\mu}$(A) and $F_{M\mu}$(C) at the
wavelength of \MgII~(Figs.~\ref{fig:fmfmuAB}
and~\ref{fig:fmfmuBC}). Because it is microlensed in both A \& C, the
very broad component VBC in \MgII~should be emitted in a very compact
region.

We also found systematic evidence that the 2 broad components BC1 and
BC2 used to decompose the BELs are differently micro-lensed in image
C. We measured $\mu_C({\rm BC2}) < \mu_C({\rm {BC1}}) < 1$ for
\MgII, \hbeta~and \halpha. This demonstrates that the broadest
component BC2 is more strongly de-amplified than the narrowest
broad-line BC1, indicating that BC2 is more compact than
BC1. Additionally, we found with {\it MCD} that $\mu_C({\rm NC1})=1$
for the narrow \hbeta~emission. Consistently, this narrow component at
$z=0.654$ is seen in $F_M$(C) for both \hbeta~and \hgamma. However,
the \hbeta~and \hgamma~profiles in $F_M$(C) cannot be fitted by a
single narrow Gaussian profile at $z=0.654$. Especially, the foot of
the line is too large, likely due to the differential micro-lensing of
the broad component of the line. Contrary to what is observed for the
Balmer lines, there is no narrow (not micro-lensed) \MgII~emission at
$z=0.654$, but a narrow {\it absorption} doublet
(Sect.~\ref{subsec:absorption}). On the other hand, a symmetric
\MgII~emission line is observed at the systemic redshift in
$F_M$(C). A Gaussian fit of this profile shows that its $FWHM$ is $\sim
2500$\,km s$^{-1}$. However, we cannot assess whether this component is
kinematically isolated (i.e. ``narrow'' \MgII~emission doublet with
$FWHM$$\sim 1000$\,km s$^{-1}$) or not (i.e. associated with differential
microlensing of the broad \MgII\ line).

A more speculative result concerns the microlensing of the Balmer
lines in image A. We observe with the $F\mu$ decomposition that only
the wings of the \hbeta~and \hgamma~profile are affected by some
micro-lensing (Fig.~\ref{fig:fmfmuAB}). On the other hand, the {\it
  MCD} analysis reveals that $\mu_A({\rm {BC1}})>1$ while $\mu_A({\rm
  {BC2}})<1$. This apparent difference of micro-lensing regime between
BC1 and BC2 (i.e. amplification of BC1 and de-amplification of BC2) is
likely not a real effect but is due to the MCD method which assumes
that Gaussian profiles describing BC1 and BC2 are microlensed as a
whole{\footnote{This assumption is not valid for image A as suggested a
    priori by the evidence for a partial micro-lensing of the BELs in
    that image while micro-lensing de-amplify nearly completely the
    BELs in C.}}. Such a measurement is indeed compatible with the
central part of the line profile being less micro-lensed than the
wings, such that an excess of flux is measured in the core of the line
(i.e. BC1), mimicking a micro-{\it amplification} of the
latter. Although only the wings of \hgamma~ are observed in
$F_{M\mu}$(A), the profile is not an exact replica of
\hbeta. Especially, the reddest fraction of \hgamma~is narrower than
its homologue in \hbeta. Additionally, it coincides with possible
\OIII\,$\lambda$4363 emission at $z=0.654$. This would reveal that at
least a fraction{\footnote{Because narrow emission -possibly
    associated with \hgamma- is also present at this wavelength in
    $F_{M}$(A) and $F_{M}$(C), it might be that only a fraction of
    \OIII\,$\lambda$4363 is micro-lensed.}} of \OIII\,$\lambda$4363 is
coming from a region more compact than the other narrow emission lines
(e.g. \hbeta, \OIIId, \NeV) and even as compact as the inner part of
the BLR. All these results underline that differential micro-lensing
of the BEL is important in image A but cannot be fully understood with
our phenomenological approach. Nevertheless, we interestingly notice
that micro-lensing of only the wings of the line profile is predicted
in Abajas et al. (\cite{ABA02}) for an outflowing broad emission
region. On the other hand, the existence of a compact
\OIII\,$\lambda$4363 emission is supported by the work of Nagao et
al. (\cite{NAG01}) who show that a significant fraction of
\OIII\,$\lambda$4363 is likely formed closer to the continuum than the
other main NELs. Because the critical density of \NeV~is only two
times smaller than for \OIII\,$\lambda$4363, one would expect \NeV~to
be formed in a region similar to \OIII\,$\lambda$4363 (Nagao et
al. \cite{NAG01b}) and thus be partially micro-lensed, which is not
observed.

In summary, we have brought several pieces of evidence that the Balmer
and \MgII~BLRs are different. Especially, we have found that there is
a very broad {\it compact} component in \MgII, not present in the
Balmer lines. The narrow emission found in \hbeta~and \hgamma~at the
same redshift as the other NELs (i.e. \OIIId, \NeV) is absent for
\MgII~ where instead narrow absorption lines are observed. The
micro-lensing of the BELs also confirms that the size of the emission
region is anti-correlated with the $FWHM$ of the line. Finally, it seems
that the micro-lensing of the BELs in image A resolves the BLR in
velocity. However, in order to discriminate between different BLR
emission models, multi-epoch data as well as a realistic modelling of
the BLR (e.g. Abajas et al.~\cite{ABA02}, Lewis \& Ibata~\cite{LEW04})
are necessary. This will be investigated in a future paper.

\subsection{Micro-lensing of the pseudo-continuum}
\label{sec:microFeII}

\begin{figure}
\centering
\includegraphics[angle = -90, width=0.95\columnwidth]{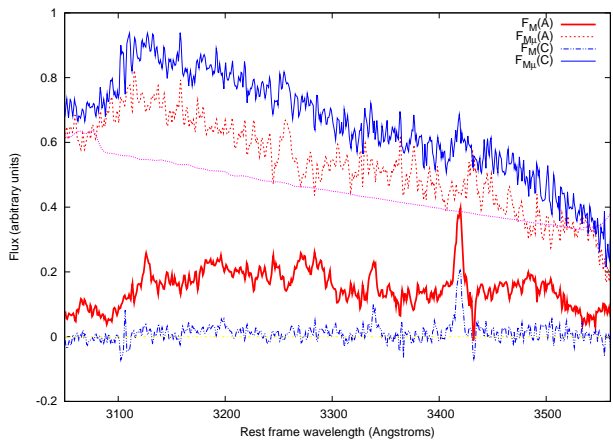}
\caption{Zoom on the $3080 < \lambda < 3540$\,\AA~ range (rest frame)
  for $F_{M\mu}$(A) (dashed red) and $F_{M\mu}$(C) (thin solid blue) and
  $F_{M}$(A) (thick solid red) and $F_{M}$(C) (dashed-dotted blue). The pseudo
  continuum model of $F_{M\mu}$ is overplotted in short dashed
  pink lines. The emission in $F_{M}$(A) is not compatible with 0 and is
  likely \FeII~emission. However, there is a conspicuous excess of
  flux in $F_{M\mu}$(A) w.r.t. the model (dotted pink line) which might also
  be due to \FeII. For image C, all the emission in that range is
  micro-lensed. }

\label{fig:Fe_excess}
\end{figure}

The $F\mu$ decomposition of the spectra (Fig.~\ref{fig:fmfmuAB}
\&~\ref{fig:fmfmuBC}) shows that a significant fraction of the
\FeII~is microlensed in image C but not in image A. This is confirmed
by the {\it MCD} analysis for which we measure similar flux ratios in
the \FeII$_{\rm opt}$ (and possibly \FeII$_{\rm UV}$) and in the
BELs. This indicates that \FeII~is emitted in a region similar to the
BLR. This is not surprising since \FeII~emission lines are generally
broadened like the broadest part of Balmer lines (e.g. Sulentic et
al. ~\cite{SUL00}). Although our result does not prohibit that
\FeII~emission arises in an accretion disk as generally believed
(e.g. Collin-Souffrin et al.~\cite{COL80}, Zhang et al.~\cite{ZHA06}),
it does not favour a scenario where {\it all} the \FeII~is emitted in the
very inner part of the disk. Nevertheless, we will show hereafter that
we have a hint that a fraction of the \FeII~emission is emitted in a
more compact region.

We can see in $F_{M\mu}$(A) that there is an excess of flux just
blueward of the \hbeta~emission (i.e. rest frame range
4630-4800\,\AA). Additionally, emission is still observed at the
same wavelength in $F_{M}$(A). When $F_{M}$(A) (redward of \hgamma) is
fitted with the \FeII$_{\rm opt}$ template, we find a lack of emission
w.r.t. the template in the range 4630-4800\,\AA. This suggests
that the excess of flux in $F_{M\mu}$(A) corresponds to \FeII~emission
missing in $F_{M}$(A). In the range 4630-4800\,\AA, \FeII~is
mainly produced by the four following multiplets: [\FeII]~4F,
[\FeII]~20F, \FeII~50 and \FeII~43 (V\'eron-Cetty et
al.~\cite{VER04}). If the micro-lensed \FeII~is associated with a
particular \FeII~transition, one would expect to observe other
components of the multiplet associated with that transition elsewhere
in $F_{M\mu}$(A). Among the 4 mentioned multiplets, \FeII~50 and
\FeII~43 only emit in the range 4630-4800\,\AA, but [\FeII]~4F is
also emitting at $\lambda$~4889\,\AA~and [\FeII]~20F is also
emitting at \lll 4874, 4905\,\AA~and around \OIIId. Since these other
components of the multiplet are significantly blended with the
\OIII~and \hbeta~emissions, we indeed cannot identify which are the
micro-lensed \FeII~transition(s). In conclusion, the present data
indicate that at least a fraction of \FeII~ in the range
4630-4800\,\AA~is emitted in a compact region, possibly as compact
as the VBC of \MgII. 

Similarly to what is observed in the range 4630-4800\,\AA, an
excess of flux is also observed in the rest-frame range
3080-3540\,\AA~(where no \FeII~templates exist). Indeed, we can
see in Fig.~\ref{fig:Fe_excess} that a fraction of the emission in
that range is microlensed in A \& C while another fraction is
micro-lensed only in image C. Although it is known that the emission
in that range is mainly due to \FeII, we cannot demonstrate that the
micro-lensing behaviour we observe is associated with two different
\FeII~emitting regions (as for the range 4630-4800\,\AA) or
not. If not associated with \FeII, we have no credible emission
candidate{\footnote{Due to the steep flux increase observed in the
    range $\sim$ 3040-3120\,\AA, the 3080-3540\,\AA~range was
    imperfectly modelled with our Balmer continuum model, suggesting
    that the microlensed flux in $F_{M\mu}$(A) is not related to the
    Balmer continuum.}} for the fraction of the emission (in the range
3080-3540\,\AA) micro-lensed in A \& C. This open question needs
careful identification of the \FeII~emission -and absorption- lines
in the range 3080-3540\,\AA. This is beyond the scope of the
present paper.

\section {Additional results}
\label{sec:complementary}

\subsection{The NEL extension}

The spectrum extraction of A, B and C has revealed that the spatial
profile of the \OIIId~region is significantly different from the one
of the underlying continuum especially for images A and C. Indeed,
after extraction, significant residuals are present around both images
A \& C while nothing comparable is observed around image B
(Sect.~\ref{subsec:extract}). This indicates that the flux in the NEL
is partly resolved along the slit direction. On the other hand, we did
not find a significant offset (i.e. offset $<$ 1 pixel) of the
centroids of the NEL images w.r.t. to the centroid of the continuum
emission. Using a Singular Isothermal Ellipsoid + external shear
(SIE+$\gamma$) to model the lensing galaxy (Paper I), we calculate
that the absence of such an offset implies an upper limit of roughly
$\sim 30 h^{-1}$\,pc on the off-centering of the \OIII~emission
w.r.t. the continuum emission. Once we have noticed the absence of an
off-centering of the \OIIId~emission, we can safely impose a limit on
its size. Indeed, using the SIE+$\gamma$ lens model, we find that
images A, B and C are merging for a source size radius{\footnote{We
    should notice that this estimate is model dependent. If
    e.g. multipoles are added to the lens model (Paper II), one finds
    that images A, B and C are merging for emission regions
    $>36h^{-1}$\,pc. }} $\geq 110 h^{-1}$\,pc. This size is compatible
with the typical size of the NLR ($\sim 1$\,kpc; Bennert et
al.~\cite{BEN02}). Stronger constraints could be derived by combining
integral field spectroscopy of J1131 with a proper method to estimate
uncertainties due to the seeing, sampling rate and lens model
degeneracies (Yonehara ~\cite{YON06}).

\subsection{The intrinsic absorption lines}
\label{subsec:absorption}

\begin{figure}
\centering
\includegraphics[angle = -90, width=0.95\columnwidth]{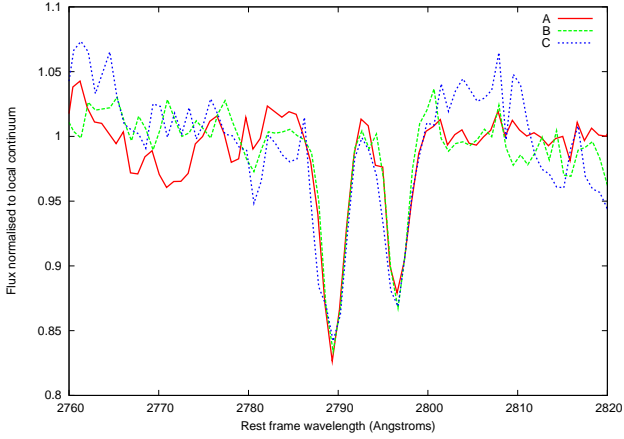}

\caption{Associated \MgII\ absorption doublet normalized to a 
    local effective continuum (i.e. continuum including the broad
  \MgII\ emission) for the 3 lensed images A, B and C. The similar
  depth in the three images indicates that the origin of the
  absorption covers both the continuum and the BLR. }

\label{fig:AAL}
\end{figure}

Another remarkable result is the presence in our spectra of a \MgII~
absorption doublet blueshifted at $z=$0.654 . According to its
velocity shift ($\Delta$v $\sim$ $-$660\,km s$^{-1}$), this absorption system
can be classified as an ``associated absorption line'' (AALs; see
e.g. Hamann \& Sabra~\cite{HAM03}).

These lines disappear from the spectrum ratios A/B and C/B. This
indicates that the absorbed flux is proportional to the flux coming
from the continuum+BLR. This implies that the region at the origin of
the absorption must cover both the continuum and the BLRs and that,
within the uncertainties, their depths are identical in the spectra of
the three images A, B, and C. This is clearly illustrated in
Fig.~\ref{fig:AAL} where the absorption lines are normalized to a 
  local effective ``continuum'' which includes the broad \MgII\
emission line.

The relative intensities of the blue and red lines of the doublet show
evidence for partial coverage. Indeed if $I_{\rm blue}$ and $I_{\rm
  red}$ are the normalized residual intensities of the two lines of
the doublet and $\tau$ the optical depth, we expect $I_{\rm blue} =
e^{-\tau}$ and $I_{\rm red} = e^{-0.5 \tau}$ in the case of full
uniform coverage. The fact that $I^2_{\rm red} < I_{\rm blue}$ as
clearly seen in Fig.~\ref{fig:AAL} suggests that only a part of the
local continuum is covered by the absorption line region.  Assuming
homogeneous partial coverage (Hamann \& Sabra~\cite{HAM03}), we find
from these intensities that the covering factor of the absorption
region is around 20\% (see e.g. the relations given in Hutsem\'ekers
et al.~\cite{HUT04}). This suggests that the absorption is intrinsic
to the quasar, and exterior to the BLR.

One can see the differential microlensing at work in J1131 as a probe
of the inhomogeneities in the absorbing medium. Indeed, by lensing
more (less) strongly some regions of the source, differential
micro-lensing increases (decreases) the contribution of a fraction of
the intervening absorber to the total absorption. The nearly identical
absorption depths seen in the three images (Fig.~\ref{fig:AAL})
indicates that both the spatial distribution and the optical depth of
the absorbing clouds must be quite homogeneous over the continuum and
BLRs. This is compatible with an absorption region composed of a large
number of small absorbing clouds, their projected sizes being
significantly smaller than the continuum region.  A more quantitative
study of this effect needs modelling of the absorbing medium, of the
micro-lensing event as well as higher spectral resolution and S/N
data.

\subsection{The lensing galaxy and neighbours}
\label{subsec:lens}

\begin{figure}
\centering
\includegraphics[angle = -90, width=0.95\columnwidth]{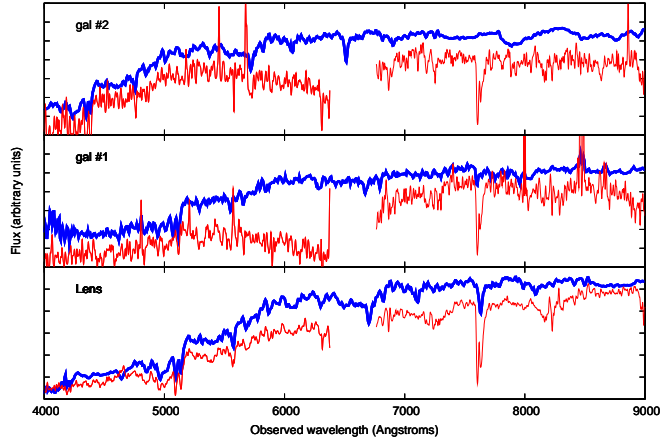}

\caption{Smoothed spectra in the blue and red ranges (thin solid
    red) of the lensing galaxy (bottom panel), gal \#1 (middle panel)
    and gal \#2 (top panel) and associated redshifted template spectra
    (thick solid blue). A flux offset has been imposed for
  legibility. Due to mis-centering of gal \#1 and \#2 in the slit,
  there is significant slit losses for $5500 < \lambda <
  6500$\,\AA~and $\lambda > 8500$\,\AA. Best redshifts are found
  to be $z$= 0.295 (Lens), $z$= 0.289 (gal \#1) and $z$= 0.105 (gal
  \#2).}

\label{fig:galaxspectra}
\end{figure}

The slit orientation \#2 provides us with a medium resolution spectrum
of the lensing galaxy and of two neighbour galaxies gal \#1 and \#2
located at 55$\arcsec$~and 95$\arcsec$~from the lens
(Sect.~\ref{subsec:extract}). We confirm the identification of the
lensing galaxy by Sluse et al. (\cite{SLU03}) as an elliptical galaxy
at $z=0.295$. Unfortunately, the S/N is too poor to derive the central
velocity dispersion. The galaxy gal\#1 is identified on deep $R$-band
imaging as a face-on spiral galaxy. There are 2 strong emission
features with an intensity ratio 2:1 observed at \lll 8461.5 \&
8488\,\AA~and a fainter one at $\lambda$\,8440\,\AA. We identify these
lines as \halphal~and \NIId~emission associated with intense star
formation. A careful look at the smoothed spectrum of the galaxy
reveals emission of \SIId, \OIIl~and \hgammal~ that definitely confirm
a redshift value of $z= 0.289 \pm 0.002$ for this galaxy. Once
smoothed with a Gaussian kernel, the spectrum of the second galaxy
(gal\#2) clearly reveals several absorption features among which Ca K
and H \lll\,3933.7, 3968.5\,\AA. The redshift of gal \#2 derived from
these absorption lines is $z=0.105 \pm 0.002$. The spectra of these
galaxies compared to the associated redshifted templates are displayed
in Fig.~\ref{fig:galaxspectra}.

Gal\#1 and gal\#2 are in the direction of the external shear (with the
convention that the shear $\gamma$ points towards or opposite
to the perturber). If we model the perturbing galaxy as a Singular
Isothermal Sphere, we can calculate the amount of shear caused by that
perturber (e.g. formula A.20. of Momcheva et al.~\cite{MOM06}). Using
a conservatively large value $\sigma=$ 300\,km s$^{-1}$ for the velocity
dispersion, we estimate that the amount of shear caused by gal\#1 and
gal\#2 is only $\gamma\sim 0.011$ for each galaxy\footnote{Although
  the angular separation from the lens is larger for gal\#2, its shear
  amplitude is as high as for gal\#1 because of its smaller
  redshift}. Even then, this is not sufficient to explain the shear
value $\gamma\sim$0.1 necessary in the lens models (e.g. Papers I \&
II). This suggests that these two perturbers might be members of more
massive galaxy groups. Interestingly, Williams et al. (\cite{WIL06})
-based on two band imaging and red sequence finding technique- found
evidence for two galaxy groups likely at $z=$0.19 and $z=$0.29 in the
field of J1131. The redshift of gal\#2 together with the measured
redshift $z=0.1035$ and $z=0.1006$ of two other field galaxies
obtained during the Las Campanas redshift survey (Shectman et
al.~\cite{SHE96}) suggest a redshift $z \sim 0.1$ for the foreground
galaxy group. These two possible groups might play a role in the
modelling of that system. This will be further investigated elsewhere.

\section{Caveats}
\label{sec:caveats}

In this section, we first further discuss the use of \OIII~as a
reasonable estimate of the macro-lens model flux ratios. Second we estimate
whether the results might be affected by the time delay between
A-B-C. Third, we briefly address the question of the effect of
intrinsic quasar absorber that might modify the BEL profile
independently of any micro-lensing.

\subsection{Macro-amplification ratios}
\label{subsec:macroratio}
Although it seems reasonable to use the flux ratios in \OIII~as an
estimate of the macro-model flux ratio, one may ask how robust our
conclusions are with respect to that assumption. For this purpose, we can
compare the flux ratios in \OIII~to those expected by a simple smooth
lens model. Using the SIE+$\gamma$ model derived in Paper I as a
fiducial model, we find $F_A/F_B= 1.65$ and $F_C/F_B=0.9$ while the
values measured in \OIII~are $F_A/F_B= 1.97$ and
$F_C/F_B=1.33$. Although there are significant differences likely due
to an imperfect modelling of the lens potential, the values measured
in \OIII~are more similar to the model than those measured in the BEL
or in the continuum. Therefore, using the fiducial model estimates of
the macro model flux ratios will not change our conclusions. Indeed,
using the macro model values instead of the \OIII~flux ratios as a
proxy of the macro-amplification ratios would only change the
amplitude of micro-lensing in A and C but nothing else.

One may ask whether the \OIII~emitting region is large enough to be
unaffected by micro-lensing. Indeed, using equation~\ref{equ:micro},
one find that a region larger than 100\,pc can be significantly
microlensed by substructures more massive than typically
$5\,10^7M_\odot$. Such a massive substructure is unlikely to affect
image A. Indeed, if present, it should significantly modify the BEL
while we have evidence that the latter is only marginally de-amplified
compared to the continuum. On the other hand, milli-lensing of image C
is more likely since we observe a significant de-amplification of a
large fraction of the BLR. Consequently, we may hypothesize that the
NLR in C is partly demagnified. The only consequence of such an
assumption would be that the effect of micro-lensing of the continuum
and the BELs in C are stronger than estimated. Alternatively, {\it
  magnification} (instead of {\it de-magnification}) of the NLR seems
incompatible with the data. Indeed, if the NLR is magnified by
substructures, one expects that the BLR and the continuum will be more
strongly affected, which is not observed. Thus, only an ad-hoc
scenario in which the NLR is magnified by a massive substructure
(e.g. a dwarf galaxy) and the BELs and continuum are de-magnified by
stars (e.g. the stars in the dwarf) would be compatible with the data.

\subsection{Bias due to the time delay}
Recently, Morgan et al.(\cite{MOR06}) argued that the time-delay
between A and B (C) is around 15 days. One may thus imagine that time
variability between A and B (C) may play a role in the interpretation
of our measurements. However, over the nearly 2 years of monitoring
performed by Morgan et al., intrinsic variability never exceeded 0.05
mag over a period of 15 days. Intrinsic variability may thus be
considered as an additional source of noise affecting our measurements
at a level of typically 5\%. This is not only true for the continuum
but also for the BEL that is mainly formed through photoionisation
processes which imply that the BEL flux variations respond linearly
to the flux variations of the continuum (e.g. Peterson et
al.~\cite{PET85}). Consequently, the time-delay between A and B (C)
should only marginally affect our results.

\subsection{Bias due to intrinsic absorbers}
Green (\cite{GRE06}) recently investigated how the small line of sight
differences existing between lensed images may modify the observed
emission line profiles. Green suggests that some warm absorber outflow
located close to the QSO continuum might modify the BEL line profile
since it would be present along the line of sight of one image but not
of the other. However, if intrinsic absorbers modify the BEL in A-B-C
it is unrealistic that they would affect the whole emission
profile. Partial ``deformation'' of the broad line profiles due to a
warm absorber might bias the $F\mu$ decomposition of the BELs
(Sect.~\ref{subsec:exploreFmu}), because the method assumes a similar
intrinsic profile in each component. The similarity between the B and
C line profiles suggests that BEL differences due to line-of-sight
effects are not significant in these images. If present, this effect
likely occurs in image A for which the \hbeta~line profile appears
more asymmetric. However, \hbeta~absorbers are very rare: up to now,
broad absorption lines in \hbeta~have been observed only in 4 systems
(Hall~\cite{HAL06}). Long term spectroscopic monitoring will shed
light on such an effect if present at all.

\section{Conclusions}
\label{sec:conclusions}

This paper is devoted to a thorough analysis of the long slit spectra
obtained for the gravitationally lensed quasar J1131-1231. The spectra
of the three brightest images A-B-C enabled us to estimate the flux
ratios in different emitting regions (namely the continuum, the broad
line and the narrow line emitting regions) and unveil the
micro-lensing effects occurring in that system. Evidence for
differential micro-lensing of the broad emission lines enabled us to
perform a phenomenological study of the structure of the Broad Line
emitting Region (BLR).

Due to the better resolving power of the present spectra, new
redshift estimates of the source and of the lensing galaxy have been
performed. Based on the \MgII~emission, a systemic redshift
$z=$0.657$\pm$0.001 has been deduced. The narrow emission lines
(except \OIIl; $z=0.656\pm0.001$) are blueshifted w.r.t. the systemic
redshift. A redshift $z=0.654\pm0.001$ is measured for those lines. On
the other hand, we confirm the redshift of the lensing galaxy to be
$z=0.295\pm0.001$. Finally, we have estimated the redshifts of the 2
galaxies gal \#1 and \#2 located at resp. 55$\arcsec$ and 95$\arcsec$
from the lens to be resp. $0.289 \pm 0.002$ and $0.105\pm0.002$.

The results derived from the micro-lensing analysis and its
implications for the quasar structure are summarized below:

\begin{enumerate}

\item Different flux ratios were derived for the continuum, the Broad
  Emission Lines (BELs) and the Narrow Emission Lines (NELs),
  indicating that micro-lensing is at work in at least 2 images. The
  simplest scenario explaining our observations consists in
  micro-lensing de-amplification of images A and C. Additionally, we
  found evidence for differential micro-lensing of the BLR
  {\footnote{During the referee process, Sugai et al. (\cite{SUG07})
      published an analysis of IFS data of J1131 in the
      \hbeta\,-\OIIId\, range. Although obtained at a different epoch
      (Feb. 2005), these data display emission line flux ratios quite
      similar to those presented here and interpreted as micro-lensing
      of images A and C. Similarly to our results, these data also
      suggest differential micro-lensing of the BLR as well as
      spatially resolved \OIII\, emission.}}. Since a larger fraction
  of the BELs is micro-lensed in image C, the Einstein radius of the
  micro-lens is likely larger in C than in A. Using the relation
  between the size of the BLR and the luminosity of the QSO (Kaspi et
  al.~\cite{KAS05}, Bentz et al.~\cite{BEN06}), we found that a
  micro-lens of a few solar masses is sufficient to affect the whole
  BLR.

\item We found that the \MgII~and the Balmer emission lines cannot be
  represented with the same velocity components. Especially, the very
  broad component of the \MgII~emission is not present in the Balmer
  lines. Since it is micro-lensed in image C and A, this component
  likely comes from a very compact region.

\item The differential micro-lensing of the BELs confirms that the
  size of an emission line region is anti-correlated with the $FWHM$ of
  the corresponding line component.

\item We have an indication that differential micro-lensing in image A
  resolves the BLR in velocity. Monitoring of such an effect
  combined with modelling of the BLR offers interesting avenues to
  probe the BLR structure and kinematic.

\item We found that the narrow emissions observed in the Balmer lines
  have the same characteristics as the \OIIId~narrow emission, namely
  same width, redshift and magnification ratio. This argues in favour
  of a common emitting region. On the other hand, we rule out the
  existence of a similar narrow emission for \MgII.

\item An intriguing narrow micro-lensed emission incidentally coincides
  with the \OIII\,$\lambda$4363 emission, suggesting a very compact
  emission region for that line. The robustness of this result is
  however questionable. Indeed, this emission might also be produced by
  the differential micro-lensing of \hgamma. Additionally, one would
  expect to see a similar behaviour for e.g. \NeV~emission, which is
  not observed. This result clearly needs further investigation. 

\item The \OIIId~narrow emission lines are partly spatially
  resolved. Neglecting seeing, resolution effects, and lens model
  degeneracies, we argue that it provides us with a lower limit on the
  size of the NLR of $110 h^{-1}$\,pc, fully consistent with typical
  NLR size.

\item We have firm evidence that a large fraction of the near-UV and
  optical \FeII~emission arises in the outer parts of the BLR. We have
  however shown that a smaller fraction of the \FeII~is emitted in the
  inner parts of BLR. Compact emission from \FeII~is very likely
  identified in the rest frame range 4630-4800\,\AA. Micro-lensed
  pseudo continuum emission is present in the range
  3080-3540\,\AA. This emission is likely associated with \FeII.

\item An associated absorption line doublet was observed in \MgII~at
  the same redshift as the NELs. We have found that the absorbing
  medium is intrinsic to the QSO and covers both the continuum and BLR
  emission, with a covering factor of $\sim$20\%. Additionally, using
  differential micro-lensing as a probe of the inhomogeneities in the
  absorbing medium, we have shown that both the spatial distribution
  and the optical depth of the absorbing clouds are homogeneous over
  the BLR and the continuum.

\end{enumerate}

Although our phenomenological description of the microlensing of the
BELs is imperfect in several aspects, it nicely illustrates the wealth
of information one can retrieve from spectroscopic observations
of micro-lensed QSOs. Clearly, more detailed models as well as
coordinated spectroscopic (integral field) monitoring of lensed
quasars where micro-lensing takes place will improve our understanding
of both the micro-lensing phenomenon and of the QSO structure.
 
\begin{acknowledgements}

  We thank M. Vestergaard for kindly providing us her UV
  \FeII~template and Monique Joly for useful discussions about
  \FeII~emission in quasars. We also acknowledge Chris Lidman for his
  tips on the NIR spectral reduction and wavelength calibration. DS is
  supported by the Swiss National Science Foundation. JFC and JS
  acknowledge support from the ESA PRODEX Programme ``HST Imaging of
  Field and Gravitationally Lensed Quasars'' and the Belgian
  Federal Science Policy Office for their support.

\end{acknowledgements}

\end{document}